\documentclass[notitlepage]{revtex4-1}

\usepackage[utf8]{inputenc}
\usepackage[colorlinks=true,citecolor=blue,linkcolor=blue]{hyperref}
\usepackage[normalem]{ulem}
\usepackage{url}
\usepackage{graphicx,wrapfig,float,slashed,cancel}
\usepackage{amsmath,amssymb,epsfig,xcolor,stmaryrd}
\usepackage{bm}
\usepackage{enumitem}

\usepackage{hhline,multirow,tabularx}  
\usepackage{graphics}
\usepackage{latexsym}
\usepackage{rotating}
\usepackage{subfigure}
\usepackage{color}
\usepackage{blindtext}
\usepackage{lipsum}
\usepackage{physics}

\begin{document}


\title{Determination of the generalized parton distributions through the analysis of the world electron scattering data considering two-photon exchange corrections}

\author{Hadi Hashamipour$^{a}$}
\email{h\_hashamipour@ipm.ir}

\author{Muhammad Goharipour$^{b,a}$}
\email{muhammad.goharipour@ipm.ir}
\thanks{Corresponding author}

\author{K.~Azizi$^{b,c,a}$}
\email{kazem.azizi@ut.ac.ir}

\author{S.V. Goloskokov$^{d}$}
\email{goloskkv@theor.jinr.ru}

\affiliation{
$^{a}$School of Particles and Accelerators, Institute for Research in Fundamental Sciences (IPM), P.O. Box 19395-5746, Tehran, Iran\\
$^{b}$Department of Physics, University of Tehran, North Karegar Avenue, Tehran 14395-547, Iran\\
$^{c}$Department of Physics, Do\u gu\c s University, Ac{\i}badem-Kad{\i}k\"oy, 34722 Istanbul, Turkey\\
$^{d}$Bogoliubov Laboratory of Theoretical Physics, Joint Institute for Nuclear Research, Dubna 141980, Moscow region, Russia}

\date{\today}

\preprint{}

\begin{abstract}

We determine the valence generalized parton distributions (GPDs) $ H_v^q $ and $ E_v^q $ with their uncertainties at zero skewness by performing a $ \chi^2 $ analysis of the world electron scattering data considering two-photon exchange corrections. The data include a wide and updated range of the electric and
magnetic form factors (FFs) of the proton and neutron. As a result, we find that there are not enough constraints on GPDs $ E_v^q $ from FFs data solely though $ H_v^q $ are well constrained.
By including the new data of the charge and magnetic radius of the nucleon in the analysis, we show that they put new constraints on the final GPDs, especially on $ E_v^q $. Moreover, we calculate the gravitational FF $ M_2 $ and the total angular momentum $ J^q $ using the extracted GPDs and compare them with the FFs obtained from the light-cone QCD sum rules (LCSR) and Lattice QCD. We show that our results are interestingly in a good consistency with the pure theoretical predictions.

\end{abstract}


\maketitle

\renewcommand{\thefootnote}{\#\arabic{footnote}}
\setcounter{footnote}{0}

\section{Introduction}\label{sec:one} 

It is well known now that the three-dimensional (3D) description of hadrons can be accessed by studying GPDs~\cite{Muller:1994ses,Radyushkin:1996nd,Ji:1996nm,Ji:1996ek,Burkardt:2000za}, which are physically related to the parton distribution functions 
(PDFs)~\cite{NNPDF:2017mvq,Hou:2019efy,Bailey:2020ooq}, e.g. with the help of double 
distribution representation~\cite{Muller:1994ses,Radyushkin:1998bz}.

The exclusive processes at large photon virtuality $Q^2$ such as the deeply virtual compton scattering (DVCS)~\cite{Ji:1996ek,Radyushkin:1997ki,Collins:1998be}, 
deeply virtual meson production (DVMP)~\cite{Goeke:2001tz,Vanderhaeghen:1999xj,Goloskokov:2005sd,Goloskokov:2006hr,Goloskokov:2007nt}, and wide-angle Compton scattering~\cite{Radyushkin:1998rt,Diehl:1998kh} 
factorize into the hard subprocess, that can be calculated perturbatively, and  
the soft part determined by GPDs~\cite{Ji:1996nm,Radyushkin:1997ki,Radyushkin:1996ru,Collins:1996fb}. Note that in most processes, GPDs contribute in integrated form that prevents direct extraction
of GPDs quantities from the experiment.

One advantage of GPDs over PDFs is that they provide quantitative information on 
both the longitudinal and transverse distributions of partons inside the 
nucleon. In this way, the structure of the nucleon can be investigated in more 
details using GPDs which include  more degrees of freedom. Actually, GPDs 
(whether polarized or unpolarized) are the functions of three variables; the 
fraction of momentum carried by the active quark ($ x $), the square of the 
momentum transfer in the process ($ t $), and the skewness parameter ($ \xi $),
which is a measure of non-forwardness of GPDs.  

GPDs contain the extensive information on the hadronic structure. In the forward 
limit, at zero $ t $ and $ \xi $, GPDs are reduced to usual PDFs. The important property of
GPDs is that GPDs integrated  over $ x $ are equal to the corresponding FFs~\cite{Ji:1996ek}. GPDs are also related to the charge and magnetization distributions. Information on the parton angular momenta can be found 
from Ji sum rules \cite{Ji:1996nm} using $ H^q $ and $ E^q $ GPDs.
More information on GPDs can be found e.g. in~\cite{Goeke:2001tz,Diehl:2003ny,Belitsky:2005qn}.

The complicated structure of GPDs leads to difficulties on their extraction from 
a $ \chi^2 $ analysis of the related experimental data like what was done 
for the PDFs.  As mentioned before, integrals of GPDs over $ x $ connect them with corresponding
FFs. This relation does not depend on the skewness $ \xi $.
This property gives possibility to use the reduced GPDs $ H^q $, $ \widetilde{H}^q $ and $ E^q $ at $ \xi=0 $ to extract them from FFs analyses~\cite{Diehl:2004cx,Diehl:2013xca}.
Following this procedure the analyses of the polarized GPDs $ \widetilde{H}^q $ was done in~\cite{Hashamipour:2019pgy,Hashamipour:2020kip}. 

In present study, we determine unpolarized GPDs $ H_v^q $ and $ E_v^q $ by performing a $ 
\chi^2 $ analysis of the world electron scattering data presented in 
Ref.~\cite{Ye:2017gyb} (\texttt{YAHL18}) where two-photon exchange (TPE) 
corrections have been incorporated.  We also include the data of the charge and 
magnetic radius of the nucleons in the analysis to investigate their impacts on 
the extracted GPDs.

The content of the present paper is as follows. In Sec.~\ref{sec:two}, we 
discuss the phenomenological framework which is used to extract GPDs from data 
as well as the method for calculating uncertainties. Sec.~\ref{sec:three} is 
devoted to introduce the experimental data which are included in our analysis. 
In Sec.~\ref{sec:four}, we present the results and investigate the 
goodness of fits. We also compare our results with the corresponding ones from 
other groups. We summarize our results and conclusions in Sec.~\ref{sec:five}.

%
\section{ Phenomenological framework}\label{sec:two}

As mentioned in the Introduction, GPDs are related to the nucleon FFs. In fact, FFs are
first moments of GPDs. For example, at zero skewness, the flavor FFs $ F_i^q $ ($ i=1,2 $) can be written in terms of the proton valence GPDs $ H_v^q $ and $ E_v^q $ for unpolarized quark of flavor
$ q $ as
\begin{align}
 F^q_1(t)=\int_{0}^1 dx\, H_v^q(x,t) , \nonumber \\ 
F^q_2(t)= \int_{0}^1 dx\, E_v^q(x,t). 
\label{Eq1}
\end{align}
On the other hand, the Dirac and Pauli FFs of the nucleon, $ F_1 $ and $ F_2 $, are expressed in terms of the flavor FFs $ F_i^q $ as
\begin{align}
F_i^p= e_u F_i^u + e_d F_i^d + e_s F_i^s, \nonumber \\ 
F_i^n= e_u F_i^d + e_d F_i^u + e_s F_i^s, 
\label{Eq2}
\end{align}
where $ p $ and $ n $ refer to the proton and neutron, respectively. Therefor, by measuring the nucleon FFs, one can obtain useful information about GPDs. However, the experimental results
are typically expressed in terms of the electric and magnetic Sachs form factors, $ G_E $ and $ G_M $, as
\begin{align}
\label{Eq3}
G^j_M(t) &= F^j_1(t) + F^j_2(t) \,, \nonumber \\ 
G^j_E(t) &= F^j_1(t) + \frac{t}{4m_j^2} F^j_2(t) \,,
\end{align}
where $ j=p,n $ denotes the type of nucleon. In the above equation, $ m $ is the nucleon mass and we have $G_E^p(0) = 1$, $G_E^n(0) = 0$, and $G^j_M(0) = \mu_j $, where $ \mu_j $ is the magnetic moment of the nucleon. 

According to \texttt{DK13} study~\cite{Diehl:2013xca}, the valence GPDs $ H_v^q(x,t) $ in Eq.~(\ref{Eq1}), can be expressed in terms of the ordinary valence PDFs $ q_v(x) $ as
\begin{equation}
H_v^q(x,t)= q_v(x)\exp [tf_q(x)],
\label{Eq4}
\end{equation}
where an exponential $ t $ behavior is considered. To be more precise, we use ansatz (4) to determine the $ t $-dependences of GPDs with the profile function $ f_q(x) $ which was introduced in~\cite{Diehl:2004cx} to
parameterize the $ x $-dependence of quark distribution in impact parameter
space. The profile function $ f_q(x) $ can have a simple or more flexible form. In this work, following the default analysis of \texttt{DK13}, we use the complex form 
\begin{equation}
\label{Eq5}
f_q(x)=\alpha^{\prime}_q(1-x)^3\log\frac{1}{x}+B_q(1-x)^3 + A_qx(1-x)^2,
\end{equation}
and take the valence PDFs $ q_v(x) $ from the \texttt{ABM11} set~\cite{Alekhin:2012ig} at the NLO and scale $ \mu=2 $ GeV. The physical motivation behind the profile function~(\ref{Eq5}) has been discussed in details in Ref.~\cite{Diehl:2004cx}. Note also that it leads to a better fit of the data compared with other forms as it has been shown in Refs.~\cite{Hashamipour:2019pgy,Hashamipour:2020kip}. The contribution of the strange quark in Eq.~(\ref{Eq2}) is also neglected as suggested by \texttt{DK13}.

For the valence GPDs $ E_v^q(x,t) $, one can consider the same ansatz of Eq.~(\ref{Eq4}) with a profile function which has similar form to Eq.~(\ref{Eq5}). But, for the forward limit one can not use the usual PDFs in this case. Then, we have 
\begin{equation}
E_v^q(x,t)= e_v^q(x)\exp [tg_q(x)],
\label{Eq6}
\end{equation}
with
\begin{equation}
\label{Eq7}
g_q(x)=\alpha^{\prime}_q(1-x)^3\log\frac{1}{x}+D_q(1-x)^3 + C_qx(1-x)^2.
\end{equation}
For $ e_v^q $, we consider a parametrization form similar to \texttt{DK13}
\begin{equation}
\label{Eq8}
e_v^q(x)=\kappa_q N_q x^{-\alpha_q} (1-x)^{\beta_q} (1+\gamma_q\sqrt{x}),
\end{equation}
where $ \kappa_u=1.67 $ and  $ \kappa_d=-2.03 $ are computed
from the measured magnetic moments of proton and neutron and
the normalization factor $ N_q $ can be obtained from the fact that
\begin{equation}
\label{Eq9}
\int_0^1 dx e_v^q(x)=\kappa_q.
\end{equation}

An important point about the forward limit of the GPDs and the profile functions is that they can not take arbitrary $ x $ dependence due to the fact that the densities for quarks and antiquarks
with momentum fraction $ x $ at a nominal transverse distance {\boldmath $ b $} from
the proton center must be different. To be more precise, this requirement implies a positivity condition as  
follows~\cite{Diehl:2013xca}
\begin{equation}
\label{Eq10}
\frac{[e_v^q(x)]^2}{8m^2} \leq \exp(1) \Big[\frac{g_q(x)}{f_q(x)}\Big]^3[f_q(x)-g_q(x)]\times \big\{[q_v(x)]^2-[\Delta q_v(x)]^2\big\},
\end{equation}
where $ \Delta q_v(x) $ are the valence polarized PDFs that we take them from the analysis of the \texttt{NNPDFpol1.1}~\cite{Nocera:2014gqa}. It can be concluded from the above equation that we should have $ g_q(x) < f_q(x) $ to preserve the positivity condition. In fact, as we discuss in Sec.~\ref{sec:four}, it is very difficult to find a best fit with more flexible profile functions, Eqs.~(\ref{Eq5}) and (\ref{Eq7}), and distribution $ e_v^q(x) $ in such a way that the positivity is preserved.

After describing the phenomenological framework that we use to extract GPDs from data, now it is time to introduce the minimization procedure and the method for calculating uncertainties. In order to determine the unknown parameters of the profile functions (\ref{Eq5}) and (\ref{Eq7}) as well as the distribution $ e_v^q(x) $ of Eq.~(\ref{Eq8}), we utilize the standard $ \chi^2 $ minimization method. To this aim, we use the CERN program library \texttt{MINUIT}~\cite{James:1975dr} and minimize the following function as usual,
\begin{equation}
\label{Eq11}
\chi^2= \sum_i^n \Big (\frac{{\cal E}_i-{\cal T}_i}{\delta{\cal E}_i} \Big )^2 , 
\end{equation}
where summation is performed over all data points included in the analysis. In the above equation, $ {\cal E}_i $ is the measured value of the experimental data point $ i $, while $ {\cal T}_i $ is the corresponding theoretical estimate. The experimental errors $ \delta{\cal E}_i $ associated with this measurements are calculated from systematic and statistical errors added in quadrature. In order to calculate uncertainties, we use the standard Hessian approach~\cite{Pumplin:2001ct} in which the uncertainties of desired quantity $ {\cal F} $ are calculated as 

\begin{equation}
\label{Eq11p}
[\delta {\cal F}]^2= \Delta \chi^2 \sum_{i,j} \Big(\frac{\partial {\cal F}}{\partial \eta_i} \Big)_{\eta=\hat{\eta}} H_{i,j}^{-1} \Big(\frac{\partial {\cal F}}{\partial \eta_j} \Big)_{\eta=\hat{\eta}}, 
\end{equation}
where the derivatives are taken with respect to the fitted parameters $ \{\eta_i\} $ with the optimum values $ \{\hat{\eta_i}\} $. The Hessian matrix $ H_{i,j} $ is calculated by \texttt{MINUIT} and provided at the end of fit procedure. The value of $ \Delta \chi^2 $ determines the confidence region, and we use the standard value $ \Delta \chi^2=1 $ in this work.

%
%
\section{Data selection}\label{sec:three}

One of the main processes that provide crucial information on GPDs is the elastic electron-nucleon scattering. Actually, by measuring this process one can extract the electric and magnetic FFs of the nucleon (or their ratio) which are related to GPDs as explained in the previous section. One of the methods to extract $ G_E $ and $ G_M $ from the unpolarized elastic electron-nucleon scattering is the Rosenbluth
separation which provides the separate determination of $ G_E $ and $ G_M $. However, it is believed that  the effects of TPE are substantial in this extraction method. There is another method to extract electromagnetic FFs in which the correlation between the polarizations of the beam electron and the proton target (or the scattered proton) is used. An advantage of this method is that it is less sensitive to TPE effects. However, a separate determination of $ G_E $ and $ G_M $ is not possible in this method and one can only access to their ratio. 

If one considers the single-photon exchange approximation, the cross section of the electron-nucleon scattering can be written in terms of Sachs FFs as~\cite{Ye:2017gyb}
\begin{align}
\left( \frac{d\sigma}{d \Omega} \right)_{0} = \left(
\frac{d\sigma}{d\Omega} \right)_{\rm Mott} \frac{\epsilon (G_E^j)^2 + \tau
	(G_M^j)^2}{\epsilon (1+\tau)} \,,
\label{Eq12}
\end{align}
where again $ j=p,n $. Here, $(d\sigma/d\Omega)_{\rm Mott}$ is the cross section of the recoil-corrected
relativistic point particle (Mott),
\begin{align}
\label{Eq13}
\left( \frac{d\sigma}{d\Omega} \right)_{\rm Mott}
=\frac{\alpha^2}{4 E^2 \sin^4{\theta / 2}} \frac{E^\prime}{E}\cos^2{\frac{\theta}{2}} 
\,. 
\end{align}
and $\epsilon$, $\tau$ are the dimensionless kinematic variables
\begin{equation}
\label{Eq14}
\begin{split}
	\tau = \frac{Q^2}{4m_j^2} \,, \quad
	\epsilon = \left[ 1 + 2 (1 + \tau) \tan^2 \frac{\theta}{2} \right]^{-1} \,.
\end{split}
\end{equation}
In the above equations, $Q^2=-q^2$ is the negative of the momentum transfer squared $ q^2 $ to the nucleon, $\theta$ is the angle of the final state electron with respect to the incident beam direction, $E$ is the initial electron energy, $ m $ is the nucleon mass, and $E^\prime=E/[1+(2E/m)\sin^2(\theta/2)]$ is the scattered electron energy. The Born cross section of Eq.~(\ref{Eq12}) should be modified by the radiative corrections as 
\begin{align}
\label{Eq15}
d\sigma = d\sigma_{0} (1 + \delta) \,,
\end{align}
where $\delta$ depends on the kinematic variables and includes the vertex, vacuum polarization, and TPE corrections.

In this work, in order to extract GPDs $ H_v^q $ and $ E_v^q $, we use the data presented in \texttt{YAHL18} analysis~\cite{Ye:2017gyb}. These data include a wide and updated range of the world electron scattering data off both proton and neutron targets. An important advantage of these data is that the TPE corrections have also been incorporated. We use three datasets of \texttt{YAHL18}, namely, world $ R_p=\mu_p G_E^p/G_M^p $ polarization, world $ G_E^n $, and world $ G_M^n/\mu_n G_D $ data where $ G_D=(1+Q^2/\Lambda^2)^{-2} $ with $ \Lambda^2=0.71 $ GeV$ ^2 $. These datasets include 69, 38, and 33 data points, respectively. In this way, the total number of data points ($ N_{\textrm{pts.}} $) that are included in the analysis will be 140.
Overall, the data cover the  $ -t $ range from 0.00973 to 10 GeV$ ^{2} $. 

It should be noted, for the case of unpolarized electron-proton ($ ep $) scattering, the \texttt{YAHL18} analysis is also included  the original cross section data (562 and 658 data points for the world and Mainz cross sections, respectively) which cover the $ -t $ range from 0.003841 to 31.2 GeV$ ^{2} $.
An interesting idea is investigation of the impact of original cross section data on the extracted GPDs.
This can be done by performing two sets of fits of the \texttt{YAHL18} data; one considering the cross section and FFs data simultaneously and the other by removing the cross section data and considering just the FFs data. However, as mentioned at the end of Sec.~\ref{sec:two}, it is very difficult to find a best fit that preserves the positivity condition Eq.~(\ref{Eq10}). To be more precise, if one releases the program to find the optimum parameters without considering positivity condition, it leads to unphysical results given that the aim of \texttt{MINUIT} is just minimizing the $ \chi^2 $ function as far as possible. On the other hand, if one try to implement the positivity condition in the main body of the fit procedure to find the optimum distributions which preserve positivity automatically, the fit does not converge. To overcome this problem one needs to run the program repeatedly as we explain in the next section. Since including the cross section data causes the fit to become time-consuming, we abandon these data and consider only the FFs data. 

The other experimental observables related to the electromagnetic FFs that can provide crucial information about the small-$ t $ behavior of GPDs are the charge and magnetic radius of the nucleons, $ r_{jE} $ and $ r_{jM} $, where again $ j=p,n $ stands for the proton and neuron, respectively. Actually, their mean squared are defined by
\begin{align}
\left<r_{jE}^2\right>= \left.  6 \dv{G_E^j}{t} \right|_{t=0} \,, \nonumber \\ 
\left<r_{jM}^2\right>= \left.  \frac{6}{\mu_j} \dv{G_M^j}{t} \right|_{t=0},
\label{Eq16}
\end{align}
where $ \mu_j $ is the magnetic moment of the nucleon. Note that in these differentiations the terms $ \int q_v(x)f_q(x) $ and $ \int e_v^q(x) g_q(x) $ are appeared that provide more direct access to the profile functions $ f_q(x) $ and $ g_q(x) $. It is also possible to get further information about the $ e_v^q(x) $ distributions. Then, in order to investigate the impact of data of the charge and magnetic radius of the nucleons on the extracted GPDs, we also include them (4 data points) in a new analysis. To this end, we use the values quoted
in the Review of Particle Physics~\cite{ParticleDataGroup:2018ovx}
\begin{align}
\sqrt{\left<r_{pE}^2\right>} =& 0.8409 \pm 0.0004~\textrm{fm}, ~~~~~ \left<r_{nE}^2\right> = - 0.1161 \pm 0.0022~\textrm{fm}^2\,, \nonumber \\ 
\sqrt {\left<r_{pM}^2\right>} =& 0.851 \pm 0.026~\textrm{fm},   ~~~~~~~ \sqrt {\left<r_{nM}^2\right>} = 0.864^{+0.009}_{-0.008}~\textrm{fm}\,.
\label{Eq17}
\end{align}
%

%
%
\section{Results}\label{sec:four}
In this section, we present the results obtained from the $ \chi^2 $ analysis of the data introduced in the previous section. To be more precise, we first perform some analyses of the electromagnetic FFs data taken from \texttt{YAHL18}~\cite{Ye:2017gyb} to construct our base fit. Then, we include also the data of the charge and magnetic radius of the nucleons in the analysis to investigate whether they can put further constraints on the extracted GPDs.

A good approach to find the unknown parameters of the parameterized distributions in any $ \chi^2 $ analysis of the experimental data is performing a parametrization scan as it is usually used in the global analysis of PDFs~\cite{H1:2009pze}. We utilize this procedure to find the optimum values of the parameters and also overcome the positivity problem described at the end of the previous section. The main idea of this procedure is releasing the free parameters step by step and scanning the $ \chi^2 $ to see how much releasing a parameter affects the value of the $ \chi^2 $ and also the shape of the distributions obtained. One should continue adding parameters until the change in the value of $ \chi^2 $ becomes less than unity, $ \Delta\chi^2 <1 $. So, in this procedure, the parametrization form is obtained systematically. Although such a procedure may seem difficult, since it needs to run the program many times, it leads to optimum results which are physical too.

\subsection{Base fit}\label{sec:four-one}

As mentioned, we first consider only the \texttt{YAHL18} data of the electromagnetic FFs, namely $ R_p=\mu_p G_E^p/G_M^p $, $ G_E^n $, and $ G_M^n/\mu_n G_D $. By analyzing these data one can obtain some base sets of GPDs that can be used (and improved) in the next steps of investigation. 
In order to find the best fit that preserves the positivity condition, we follow two different scenarios.
\begin{itemize}
\item {\bf Scenario 1:} Since the forward limit of the GPDs $ E $, namely $ e_v^q $ of Eq.~(\ref{Eq8}), plays a crucial role
to preserve positivity as pointed out by \texttt{DK13}~\cite{Diehl:2013xca}, we start the parametrization scan by releasing parameters $ \alpha_q $ and $ \beta_q $ of Eq.~(\ref{Eq8}) and parameters $ \alpha_q^\prime $ and $ A_q $ of the profile function $ f_q(x) $ of Eq.~(\ref{Eq5}). It should be noted that we set $ \alpha_{u_v}=\alpha_{d_v} $ and fix the parameter $\alpha^{\prime}_{u_v} $ by the relation $ \alpha^{\prime}_{u_v} - \alpha^{\prime}_{d_v} = 0.1 $ GeV$ ^{-2} $ as suggested by \texttt{DK13}. Moreover, for the scale $ \mu^2 $ at which the PDFs are chosen in ansatz~(\ref{Eq4}) we also follow \texttt{DK13} and set it as $ \mu=2 $ GeV. As the next step, we fix the parameters $ \beta_{u_v} $ and $ \beta_{d_v} $ on their optimum values obtained from the previous step and release the other parameters of the profile functions $ f_q(x) $ and $ g_q(x) $ step by step. In each step we make sure that the positivity is preserved to a good extent (at least for $ x < 0.8 $) as well as the fit is converged. We reject those fits that violate strongly the positivity condition and take the ones with the lowest value of the $ \chi^2 $. The procedure is continued until the release of all parameters is checked.

\item {\bf Scenario 2:} Another approach to find the best fit that preserves the positivity condition is that we implement 
the positivity condition in the main body of the fit program. In this way, the distributions obtained preserve the positivity automatically. However, the problem is that the fit is converged hardly in this case. To solve this problem we confine ourselves to implement condition $ g_q(x) < f_q(x) $ instead of Eq.~(\ref{Eq10}). Then, we use the parametrization scan and follow the same procedure described in Scenario 1. The only difference is that we release parameters $ \beta_{u_v} $ and $ \beta_{d_v} $ from the beginning to the end of the parametrization scan. 
\end{itemize}

Following Scenario 1, we have found two sets of GPDs that preserve the positivity; one with $ B_{d_v}=0 $ and the other with 
$ B_{u_v}=0 $ which are called Set 1 and Set 2, respectively. Note that releasing these parameters (as the last parameter) leads to the positivity violation. Following Scenario 2, we have found another GPD set in which all parameters $ A_q $, $ B_q $, $ C_q $, and $ D_q $ are free (Set 3). We remind the reader that this set has an another advantage, namely the freedom of the parameters $ \beta_{u_v} $ and $ \beta_{d_v} $, that makes the error calculations of the extracted GPDs more real. Another point that should be noted is that for both scenarios (all three sets of GPDs) we considered parameters $ \gamma_q $ of Eq.~(\ref{Eq8}) to be equal to zero. In fact, we checked this issue by continuing the parametrization scan and releasing these parameters and found that they can not improve the value of $ \chi^2 $ significantly. On the other hand, taking them as free parameters leads to the positivity violation. 
Overall, one can say that the fit is more sensitive to the parameters of the up quark distribution so that releasing them leads to more decrease in the value of $ \chi^2 $ compared with the corresponding ones of the down quark.

The values of the optimum parameters of the profile functions (\ref{Eq5}) and (\ref{Eq7}), and distribution $ e_v^q(x) $ of Eq.~(\ref{Eq8}) are listed in Table~\ref{tab:par} for all three analyses described above. The parameters denoted by an asterisk ($ * $) have been fixed as explained. According to the results obtained, both scenarios lead to a large value for parameters $ \beta_{u_v} $ and $ \beta_{d_v} $. To be more precise, whether they are fixed after the first step of the parametrization scan on their optimum values or released  to the end of the parametrization scan, a large value is obtained for them. This may be strange at the first glance. However, as we will see later, this can be attributed to the fact that there are not enough constrains on $ e_v^q(x) $ from the electromagnetic FFs data solely. 
\begin{widetext}
\begin{center}

\begin{table}[th!]
\caption{The optimum parameters of the profile functions (\ref{Eq5}) and (\ref{Eq7}), and distribution $ e_v^q(x) $ of Eq.~(\ref{Eq8}) for the analyses described in Sec.~\ref{sec:four-one}. The parameters denoted by an asterisk ($ * $) have been fixed during the fit.}\label{tab:par}
\centering
{\scriptsize 
\newcolumntype{C}[1]{>{\hsize=#1\centering\arraybackslash}X}
\centering
\begin{tabularx}{0.55\linewidth}{llc*{3}{C{3cm}C{3cm}}C{3cm}C{3cm}}  \hline \hline
               & Parameter                    & Set 1            & Set 2            & Set 3                           \\ 
               \hline 
               \hline
               & $ \alpha_d^{\prime} $        & $ 0.98 \pm 0.04 $         & $ 1.02 \pm 0.04 $         & $ 0.83 \pm 0.06 $               \\
               & $ A_{u_v} $                  & $ 1.54 \pm 0.13 $         & $ 1.63 \pm 0.15 $         & $ 1.27 \pm 0.21 $               \\
               & $ B_{u_v} $                  & $ 0.15 \pm 0.08 $         & $ 0.00^* $                & $ 0.66 \pm 0.25 $               \\               
               & $ A_{d_v} $                  & $ 3.42 \pm 0.52 $         & $ 3.45 \pm 0.50 $         & $ 3.68 \pm 0.59 $               \\
               & $ B_{d_v} $                  & $ 0.00^* $                & $-0.14 \pm 0.07 $         & $ 0.46 \pm 0.22 $               \\                              
               & $ C_{u_v} $                  & $ 1.33 \pm 0.27 $         & $ 1.52 \pm 0.21 $         & $ 0.79 \pm 0.46 $               \\
               & $ D_{u_v} $                  & $-1.02 \pm 0.16 $         & $-1.15 \pm 0.13 $         & $-0.69 \pm 0.34 $               \\               
               & $ C_{d_v} $                  & $ 2.92 \pm 0.68 $         & $ 3.16 \pm 0.73 $         & $ 2.67 \pm 0.89 $               \\
               & $ D_{d_v} $                  & $-1.15 \pm 0.18 $         & $-1.24 \pm 0.20 $         & $-1.27 \pm 0.32 $               \\
               & $ \alpha_{u_v} $             & $ 0.48 \pm 0.02 $         & $ 0.46 \pm 0.03 $         & $ 0.55 \pm 0.03 $               \\               
               & $ \beta_{u_v} $              & $ 8.22^* $                & $ 8.22^* $                & $ 8.90 \pm 1.72 $               \\  
               & $ \gamma_{u_v} $             & $ 0.00^* $                & $ 0.00^* $                & $ 0.00^* $                      \\                                                          
               & $ \alpha_{d_v} $             & $ \alpha_{u_v}^*$         & $ \alpha_{u_v}^* $        & $ \alpha_{u_v}^* $              \\               
               & $ \beta_{d_v} $              & $ 9.58^* $                & $ 9.58^* $                & $14.50 \pm 4.01 $               \\  
               & $ \gamma_{d_v} $             & $ 0.00^* $                & $ 0.00^* $                & $ 0.00^* $                      \\                                                          
               \hline 
               \hline
\end{tabularx}
}
\end{table}
\end{center}
\end{widetext}

Table~\ref{tab:chi2} contains the results of three analyses of the \texttt{YAHL18} experimental data for electromagnetic FFs introduced in Sec.~\ref{sec:three} that have been performed following Scenarios 1 and 2 described above. This table includes the list of datasets used in the analysis, along with their related observables. For each dataset, the range of $ -t $ which is covered by data and the value of $ \chi^2 $ divided by the number of data points, $\chi^2$/$ N_{\textrm{pts.}} $ are presented. The last row of the table contains the values of the $\chi^2$ divided by the number of degrees of freedom ($\chi^2$/d.o.f.). As can be seen, the results obtained for all three analyses Set 1, Set 2, and Set 3 are very similar, especially for Set 1 and Set 2 that have been obtained following same scenario. However, analysis  Set 3 has led to the smaller $ \chi^2 $. Overall, the values of $\chi^2$/$ N_{\textrm{pts.}} $ for the neutron data are very well that indicates the goodness of fit for these data rather than the proton data of $R_p $.
\begin{widetext}
\begin{center}

\begin{table}[th!]
\caption{The results of three analyses of the \texttt{YAHL18} experimental data for electromagnetic FFs introduced in Sec.~\ref{sec:three} that have been performed following Scenarios 1 and 2 described in Sec.~\ref{sec:four-one}.}\label{tab:chi2}
\centering
{\scriptsize 
\newcolumntype{C}[1]{>{\hsize=#1\centering\arraybackslash}X}
\centering
\begin{tabularx}{0.63\textwidth}{llc*{3}{C{2.5cm}C{2.5cm}}} 

\hline \hline
Observable                 & -$t$ (GeV$^2$)        &  \multicolumn{3}{c}{ $\chi^2$/$ N_{\textrm{pts.}} $  }  \\
                              &    & Set 1             & Set 2                     & Set 3      \\
\hline
\hline
$R_p = \mu_p G_{E}^p / G_{M}^p$  & $0.162-8.49$   & $110 /  69$                  & $109 /  69$  & $106 /  69$      \\ 
$G_{E}^n$                        & $0.00973-3.41$ & $ 25 /  38$                  & $ 25 /  38$  & $25 /  38$        \\ 
$G_M^n/\mu_n G_D$             & $0.071-10.0$    & $44 /  33$                  & $44 /  33$  & $45 /  33$        \\
\hline
Total $\chi^2 /\mathrm{d.o.f.} $ & & $179 / 131$                & $178 / 131$ & $176 /  131$       \\
\hline
\hline
\\ 
\end{tabularx}
}
\end{table}
\end{center}
\end{widetext}

Since in the following we are going to compare our results with the corresponding ones obtained from \texttt{DK13} analysis, it is appropriate to explain more about the differences and similarities of our analysis with \texttt{DK13}. From a methodological point of view, two analysis are similar to a large extent. Although the parametrization form of the profile functions~(\ref{Eq5}) and~(\ref{Eq7}) as well as the distribution $ e_v^q(x) $ of Eq.~(\ref{Eq8}) are same, we use the parametrization scan, as mentioned before, to find the optimum values of the parameters. In our study, the positivity condition is applied more strictly compared with \texttt{DK13}. Actually, it is either checked step by step during the parametrization scan or applied directly in the main body of the fit program. For the case of data selection, as mentioned in Sec.~\ref{sec:three}, we use the data presented in \texttt{YAHL18} analysis~\cite{Ye:2017gyb} where the TPE corrections have also been incorporated. Although lots of references are the same, there are also some differences. For the case of proton data, we have used 69 $ R_p $ data, while \texttt{DK13} have used 54 data points of both $ R_p $ and $ G_M^p $ from~\cite{Arrington:2007ux} which can be considered as an older version of \texttt{YAHL18}. For the case of neutron data, we have used 38 and 33 data points of $ G_E^n $ and $ G_M^n $, respectively, while \texttt{DK13} have used 36 and 21 data points of $ G_M^n $ and $ R^n $, respectively. Overall, the new data points (including updated ones) used in our analysis are 21 data points for $ R_p $~\cite{JeffersonLabE93-049:2002asn,ResonanceSpinStructure:2006oim,Crawford:2006rz,Puckett:2011xg,Puckett:2017flj}, 12 data points for $ G_M^n $~\cite{Rock:1982gf,Lung:1992bu,Gao:1994ud,Kubon:2001rj}, and 3 data points for $ G_E^n $~\cite{Meyerhoff:1994ev,Eden:1994ji,Schlimme:2013eoz}. For the case of nucleon's radius data, we have used (see Sec.~\ref{sec:four-two}) 4 data points introduced in Eq.~\ref{Eq17}, while \texttt{DK13} have only used $ r_{nE}^2 $.

\begin{figure}[!htb]
    \centering
\includegraphics{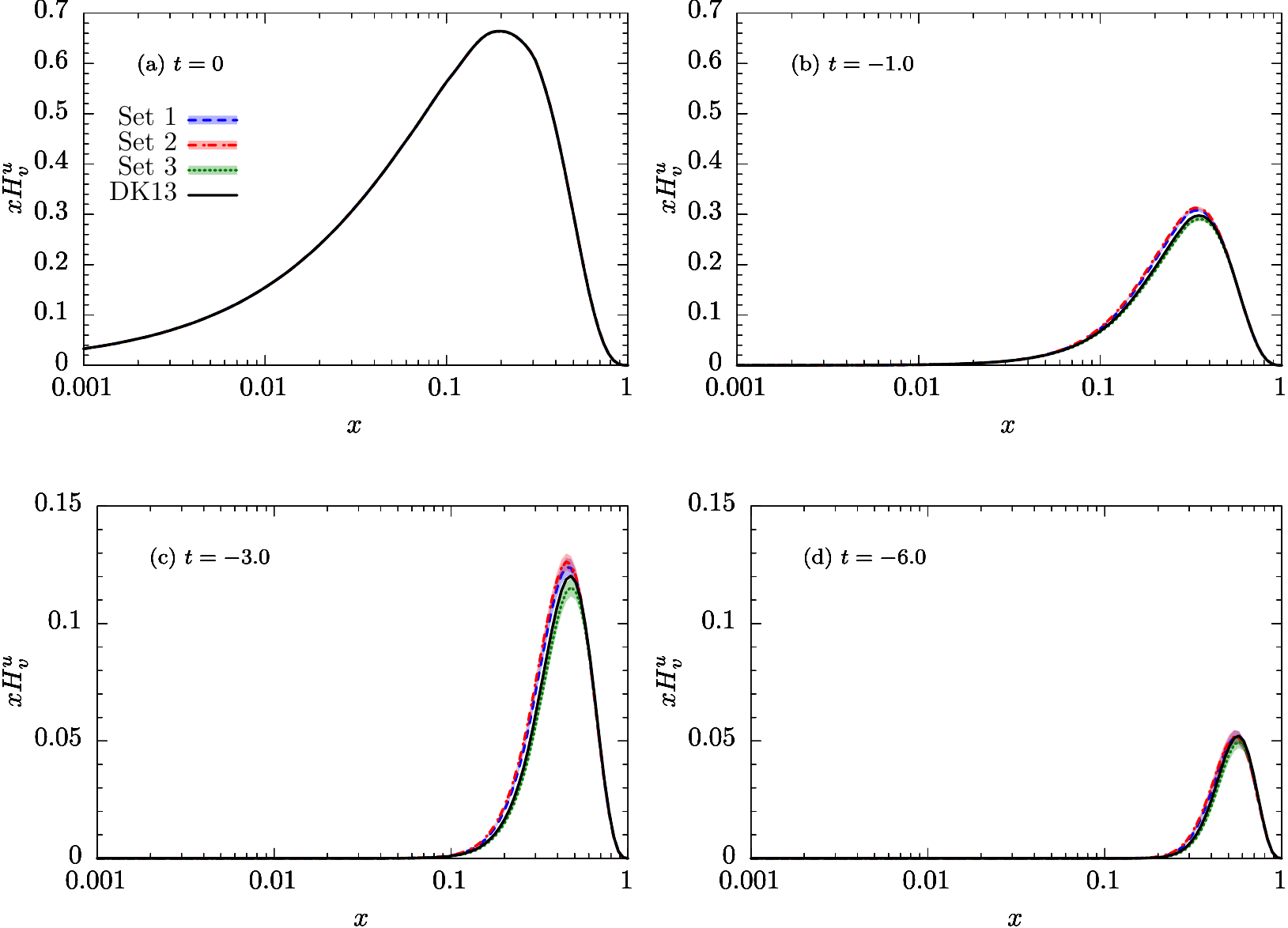}   
    \caption{A comparison between our results for GPD $ xH_v^u(x) $ obtained from three analyses
    	described in Sec. \ref{sec:four-one} and the corresponding one from the analysis
    	of DK13 \cite{Diehl:2013xca} at four $ t $ values shown in panel (a) $ t=0$, (b)
    	$t=-1$, (c) $t=-3$, and (d) $t=-6 $ GeV$ ^2 $.}
\label{fig:Hu}
\end{figure}

Figure~\ref{fig:Hu} shows a comparison between our results for GPD $ xH_v^u(x) $ obtained from three analyses described above with their uncertainties and the corresponding ones from the analysis of \texttt{DK13}~\cite{Diehl:2013xca} at four $ t $ values, $ t=0,-1,-3,-6 $ GeV$ ^2 $. As can be seen from the figure, the results are very similar, especially for Set 1 and Set 2. However, Set 3 predicts smaller distribution. It should be noted that at $ t=0 $ our results are completely consistent with \texttt{DK13} as expected, since the forward limits of GPDs $ H_v^q $ in all analyses have been taken from \texttt{ABM11}~\cite{Alekhin:2012ig} as mentioned before. It is also worth noting that the maximum position of $ H_v^u $ moves to the larger values of $ x $ with $ -t $ growing, as expected.

\begin{figure}[!htb]
    \centering
\includegraphics{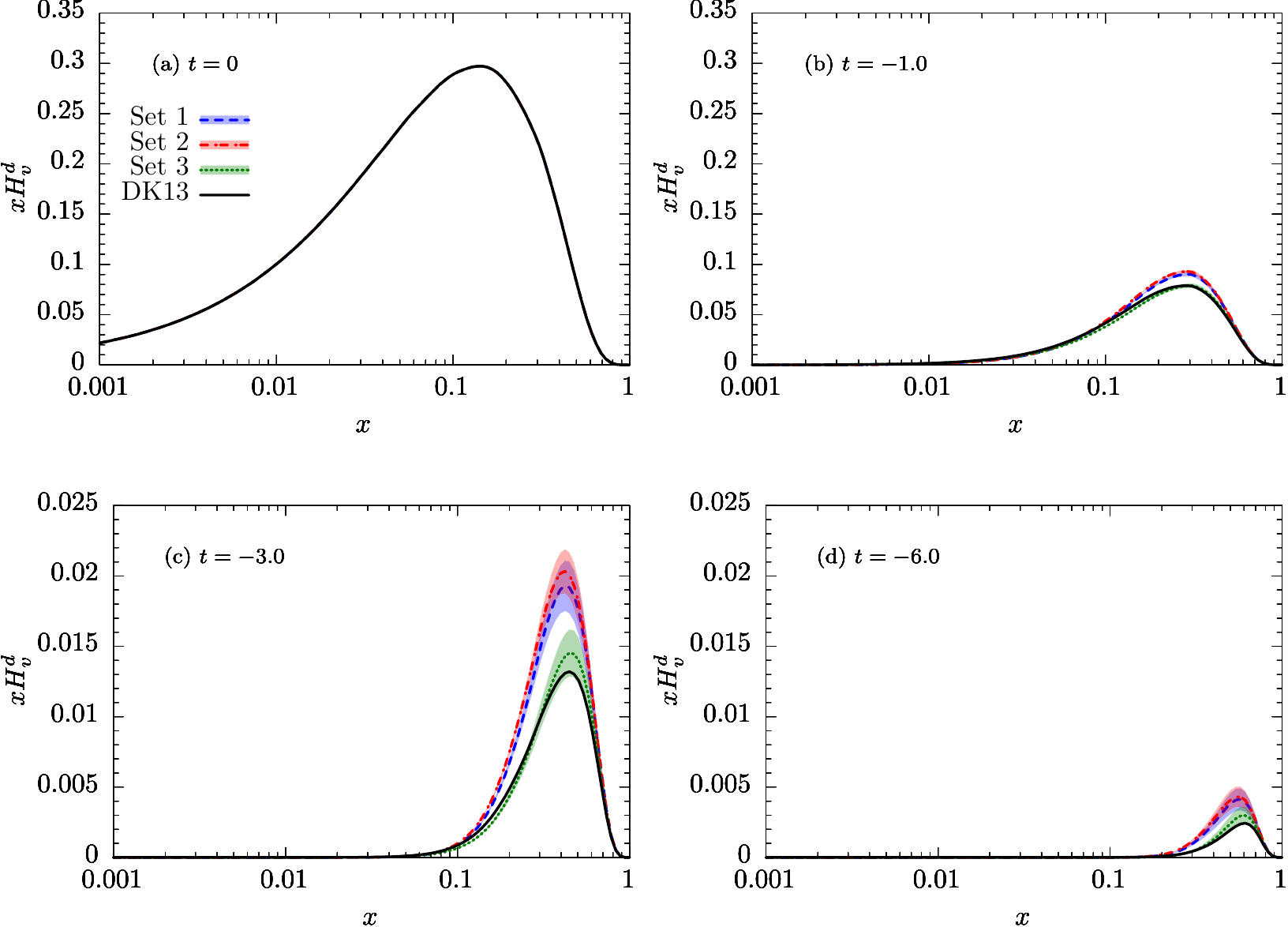}    
    \caption{Same as Fig.~\ref{fig:Hu} but for GPD $ xH_v^d(x) $. }
\label{fig:Hd}
\end{figure}

Figure~\ref{fig:Hd} shows the same results as Fig.~\ref{fig:Hu} but for GPD $ xH_v^d(x) $. In this case, Set 1 and Set 2 lead again to very similar results even with $ -t $ growing. But, there is significant difference between them and Set 3 especially when $ -t $ increases. However, Set 3 and \texttt{DK13} predict similar distributions at all values of $ -t $.  An important point that should be mentioned is that the uncertainties from PDFs in Eq.~(\ref{Eq4}) have not been considered in the error calculations of GPDs $ H_v^q $ in Figs.~\ref{fig:Hu} and~\ref{fig:Hd}. Actually, since $ H_v^q(x) $ are related to the valence PDFs $ q_v(x) $, and on the other hand global analyses of PDFs lead to precise valence parton distributions, it is expected that their uncertainties do not affect significantly the error bands of the valence GPDs $ H_v^q(x) $.

\begin{figure}[!htb]
    \centering
\includegraphics{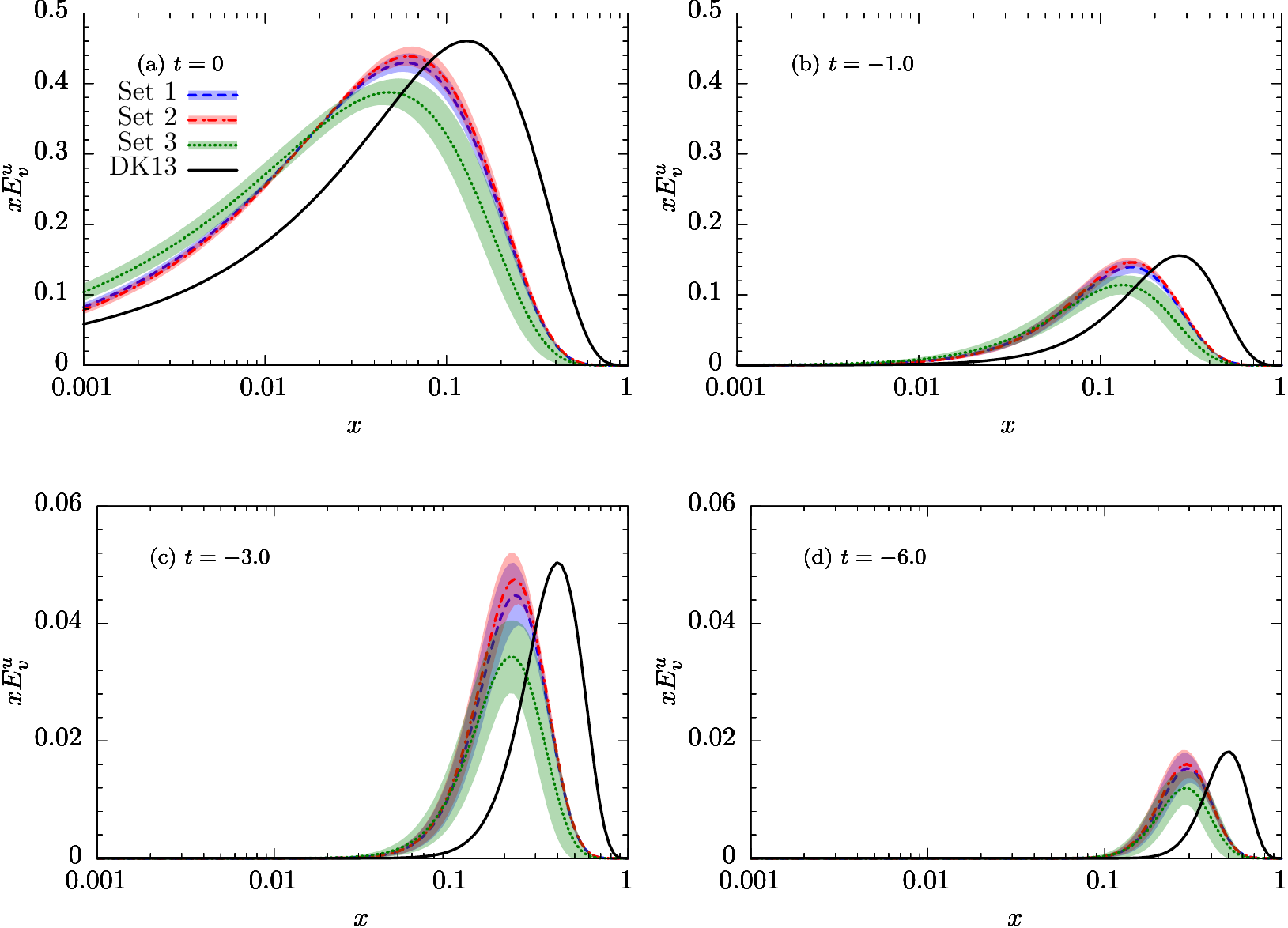}    
    \caption{Same as Fig.~\ref{fig:Hu} but for GPD $ xE_v^u(x) $. }
\label{fig:Eu}
\end{figure}

The results obtained for GPD $ xE_v^u(x) $ are shown in Fig.~\ref{fig:Eu} and compared again with the corresponding one from \texttt{DK13} at $ t=0,-1,-3,-6 $ GeV$ ^2 $. As can be seen, in this case, Set 1, Set 2, and Set 3 are almost similar, but they differ significantly with \texttt{DK13}. To be more precise, they tend to smaller $ x $ rather than \texttt{DK13}. This significant difference can be attributed to the difference in forward limits of $ E_v^u $, namely $ e_v^u(x) $, as can be concluded from the panel (a) of Fig.~\ref{fig:Eu} that shows the results with $ t=0 $.

\begin{figure}[!htb]
    \centering
\includegraphics{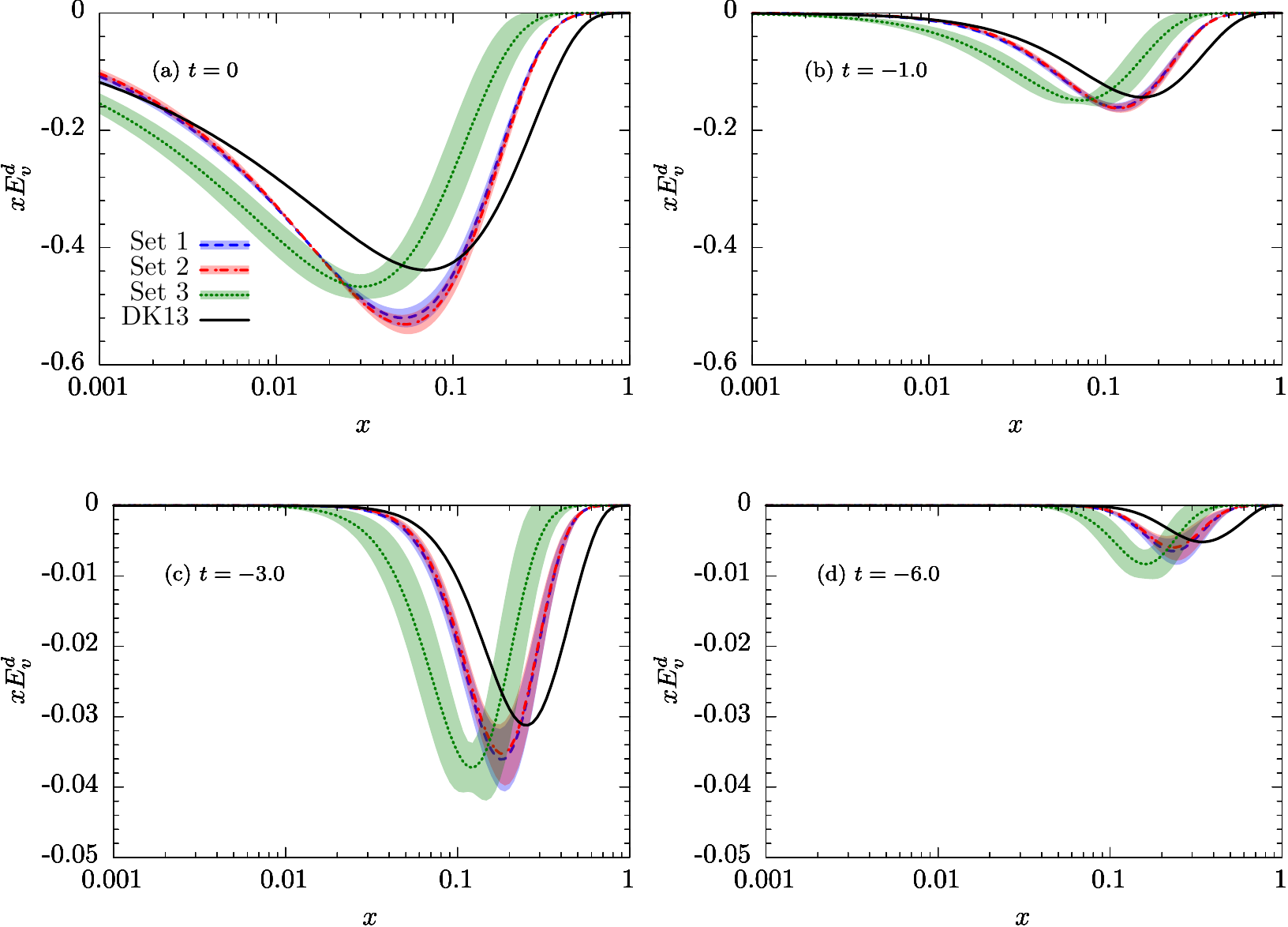}
    \caption{Same as Fig.~\ref{fig:Hu} but for GPD $ xE_v^d(x) $. }
\label{fig:Ed}
\end{figure}

Figure~\ref{fig:Ed} shows the same results as Fig.~\ref{fig:Eu} but for GPD $ xE_v^d(x) $. In this case, one can see significant difference between Set 3 and other sets that can be attributed to the difference in $ e_v^d(x) $, again, according to the panel (a) of Fig.~\ref{fig:Ed}. Moreover, Set 1 and  Set 2 are not in good agreement with \texttt{DK13}. Overall, our results tend to smaller $ x $ rather than \texttt{DK13}.

Considering all results presented in this section, we conclude that our results for GPDs $ H_v^q $ are in better agreement with \texttt{DK13} than the corresponding ones for $ E_v^q $. This difference can be due to both different data included in the analysis and different procedure to find the optimum parameters and final distributions. Note that, as mentioned before, for Set 3 we have implemented the condition 
$ g_q(x) < f_q(x) $ in the main body of the program so that the set of parameters which lead to the  violation of this condition is automatically rejected. Overall, we can say that our results show more respect for the positivity than \texttt{DK13}. In fact, we checked this issue and found that for the case of up quark our results preserve the positivity condition Eq.~(\ref{Eq10}) at all values of $ x $, while for the case of down quark it is violated just for $ x>0.8 $. However,  Eq.~(\ref{Eq10}) is violated for both up and down quarks if one considers the results of \texttt{DK13} for $ x>0.9 $ and $ x>0.8 $, respectively.

Another important conclusion can be drawn from the results obtained in this subsection is that there are not enough constraints on GPDs $ E_v^q $ from the electromagnetic FFs data. In other words,
considering such data solely does not lead to the universal GPDs of $ E_v^q $, since one can obtain different results that all preserve the positivity condition and have the same values of the $ \chi^2 $ (which means that all of them describe the data well to the same extent).

\subsection{Impact of nucleon's radius data}\label{sec:four-two}

Having a base set of GPDs in hand, now it is interesting to check how they describe the experimental data of the charge and magnetic radius of the nucleons introduced at the end of Sec.~\ref{sec:three}. To this aim, we have used the final sets of GPDs extracted in the previous subsection as well as \texttt{DK13} GPDs~\cite{Diehl:2013xca} in the calculation of Eq.~(\ref{Eq16}) and compared the results obtained with the related data in Table.~\ref{tab:rEM}. As can be seen, \texttt{DK13} predictions are in better agreement with data rather than our results. In fact, none of our sets of GPDs can predict data within the error, while \texttt{DK13} predictions of $ r_{pM}^2 $ and $ r_{nE}^2 $ are inside the error bands of the data. This can be due to the inclusion of $ r_{nE}^2 $ data in the \texttt{DK13} analysis, while our sets are obtained just by the inclusion of the electromagnetic FFs data as described in Sec.~\ref{sec:four-one}.
\begin{widetext}
\begin{center}

\begin{table}[th!]
\caption{A comparison between the theoretical calculations of the charge and magnetic radius of the nucleons obtained using the Set 1, Set 2, and Set 3 of GPDs extracted in Sec.~\ref{sec:four-one} as well as \texttt{DK13} GPDs~\cite{Diehl:2013xca} and the related data introduced at the end of Sec.~\ref{sec:three}.}\label{tab:rEM}
\centering
{\scriptsize 
\newcolumntype{C}[1]{>{\hsize=#1\centering\arraybackslash}X}
\centering
\begin{tabularx}{0.8\textwidth}{llc*{3}{C{2.5cm}C{2.5cm}}} 

\hline \hline
Observable                 & Experiment       &  \multicolumn{4}{c}{ Theory  }  \\
                              &    & Set 1             & Set 2                     & Set 3   &  \texttt{DK13}   \\
\hline
\hline
$\sqrt{<r_{pE}^2>}$              & $ 0.8409 \pm 0.0004 $         & $0.836 \pm 0.015$       & $0.836 \pm 0.013 $   & $0.837 \pm  0.017 $    & $0.839 $  \\
$\sqrt {<r_{pM}^2>}$             & $ 0.851 \pm 0.026 $           & $0.819 \pm 0.017$       & $0.819 \pm 0.017$   & $0.822  \pm 0.025 $   & $0.829 $  \\
$<r_{nE}^2>$                     & $ - 0.1161 \pm 0.0022 $       & $-0.099\pm 0.017$      & $-0.099 \pm 0.016$  & $-0.103  \pm 0.02 $  & $-0.113 $  \\
$\sqrt {<r_{nM}^2>}$             & $ 0.864^{+0.009}_{-0.008} $   & $0.827\pm 0.021$       & $0.829 \pm 0.021$    & $0.822 \pm 0.036$   & $0.848  $  \\
\hline
\hline
\end{tabularx}
}
\end{table}
\end{center}
\end{widetext}

According to the above explanations, in this subsection, we also include the data of the charge and magnetic radius of the nucleon in the analysis to see if they can put more constraints on the extracted GPDs, especially of $ E_v^q $. To this aim, we follow same scenarios explained at Sec.~\ref{sec:four-one}. The only difference is that we include also parameters $ \beta_{u_v} $ and $ \beta_{d_v} $ from the beginning to the end of the parametrization scan for the case of scenario 1, since it is expected that more constraints are provided by including the radius data.

Following Scenario 1, we find a set of GPDs with $ D_{u_v}=0 $ and $ \gamma_{d_v}=0 $ while all other parameters are free. This set is called Set 4. Note that releasing $ D_{u_v} $ leads to a strong positivity violation and releasing $ \gamma_{d_v} $ does not improve the value of total $ \chi^2 $ so that it even leads to a little enhancement of the $\chi^2 /\mathrm{d.o.f.} $ value. Following Scenario 2, we find a set of GPDs with $ D_{d_v}=0 $ and $ \gamma_{d_v}=0 $ which is called Set 5. Actually, releasing these parameters do not lead to any improvement in the $\chi^2 /\mathrm{d.o.f.} $ value.

Table~\ref{tab:par-two} compares the values of the optimum parameters of the profile functions (\ref{Eq5}) and (\ref{Eq7}), and distribution $ e_v^q(x) $ of Eq.~(\ref{Eq8}) for Set 4 and Set 5 described above. The parameters have been fixed  during the fit are denoted again by an asterisk ($ * $). As can be seen, the values of the parameters $ \beta_{u_v} $ and $ \beta_{d_v} $ are decreased seriously and become more logical after the inclusion of the radius data in the analysis. The results of these two analyses have also been compared in Table~\ref{tab:chi2-two}. It is obvious from this Table that both Set 4 and Set 5 have overall a good description of data and their $\chi^2$/$ N_{\textrm{pts.}} $ are very similar.  
As before, the goodness of fit for the neutron data is better than the proton data. For the case of charge and magnetic radius data, satisfactory results have been obtained, except for the case of $ r_{nM}^2 $.
\begin{widetext}
\begin{center}

\begin{table}[th!]
\caption{The optimum parameters of the profile functions (\ref{Eq5}) and (\ref{Eq7}), and distribution $ e_v^q(x) $ of Eq.~(\ref{Eq8}) for the analyses described in Sec.~\ref{sec:four-two}. The parameters denoted by an asterisk ($ * $) have been fixed during the fit.}\label{tab:par-two}
\centering
{\scriptsize 
\newcolumntype{C}[1]{>{\hsize=#1\centering\arraybackslash}X}
\centering
\begin{tabularx}{0.39\linewidth}{llc*{2}{C{3cm}C{3cm}}}  \hline \hline
               & Parameter                    & Set 4            & Set 5            \\
               \hline 
               \hline
               & $ \alpha_d^{\prime} $        & $ 0.941 \pm 0.063 $         & $ 0.883 \pm 0.061 $        \\
               & $ A_{u_v} $                  & $ 1.786 \pm 0.217 $         & $ 1.748 \pm 0.236 $        \\ 
               & $ B_{u_v} $                  & $ 0.249 \pm 0.221 $         & $ 0.438 \pm 0.221 $        \\ 
               & $ A_{d_v} $                  & $ 4.985 \pm 0.443 $         & $ 5.263 \pm 0.434 $        \\
               & $ B_{d_v} $                  & $ -0.125 \pm 0.219  $       & $ 0.042 \pm 0.226 $        \\
               & $ C_{u_v} $                  & $ 1.463 \pm 0.553 $         & $ 1.847 \pm 0.562 $        \\
               & $ D_{u_v} $                  & $  0.000^* $                & $ -0.725 \pm 0.447 $       \\
               & $ C_{d_v} $                  & $ 3.934 \pm 0.580 $         & $ 4.051 \pm 0.453 $        \\
               & $ D_{d_v} $                  & $-0.301 \pm 0.190 $         & $   0.000^*  $            \\
               & $ \alpha_{u_v} $             & $ 0.573 \pm 0.040 $         & $ 0.573 \pm 0.034 $        \\
               & $ \beta_{u_v} $              & $ 5.648 \pm 1.117 $         & $ 7.917 \pm 2.648 $        \\
               & $ \gamma_{u_v} $             & $ 4.150 \pm 2.917 $         & $ 0.871 \pm 0.861 $        \\
               & $ \alpha_{d_v} $             & $ \alpha_{u_v}^*$           & $ \alpha_{u_v}^* $        \\
               & $ \beta_{d_v} $              & $ 2.595 \pm 0.525 $         & $ 2.119 \pm 0.288 $        \\
               & $ \gamma_{d_v} $             & $ 0.000^* $                 & $ 0.000^* $                \\
               \hline 
               \hline
\end{tabularx}
}
\end{table}
\end{center}
\end{widetext}
%
%
\begin{widetext}
\begin{center}

\begin{table}[th!]
\caption{The results of two analyses of the experimental data introduced in Sec.~\ref{sec:three} that have been performed following Scenarios 1 and 2 described in Sec.~\ref{sec:four-two}.}\label{tab:chi2-two}
\centering
{\scriptsize 
\newcolumntype{C}[1]{>{\hsize=#1\centering\arraybackslash}X}
\centering
\begin{tabularx}{0.48\textwidth}{llc*{2}{C{3cm}C{3cm}}} 

\hline \hline
Observable                 & -$t$ (GeV$^2$)        &  \multicolumn{2}{c}{ $\chi^2$/$ N_{\textrm{pts.}} $  }  \\
                              &    & Set 4             & Set 5              \\
\hline
\hline
$R_p = \mu_p G_{E}^p / G_{M}^p$  & $0.162-8.49$      & $118.7 /  69$           & $117.1 /  69$  \\
$G_{E}^n$                        & $0.00973-3.41$    & $26.9  /  38$           & $ 27.6 /  38$  \\
$G_M^n/\mu_n G_D$                & $0.071-10.0$      & $46.9  /  33$           & $44.8 /  33 $  \\
$\sqrt{\left<r_{pE}^2\right>}$   & $ 0 $             & $0 /  1$                & $0 /  1 $  \\
$\sqrt {\left<r_{pM}^2\right>}$  & $ 0 $             & $0.6 /  1$              & $0.6 /  1 $  \\
$\left<r_{nE}^2\right>$          & $ 0 $             & $1.9 /  1$              & $1.0 /  1 $  \\
$\sqrt {\left<r_{nM}^2\right>}$  & $ 0 $             & $4.0 /  1$              & $6.9 /  1 $  \\
\hline
Total $\chi^2 /\mathrm{d.o.f.} $ & & $199 / 132$                & $198 / 132$       \\
\hline
\hline
\\ 
\end{tabularx}
}
\end{table}
\end{center}
\end{widetext}

\begin{figure}[!htb]
    \centering
\includegraphics{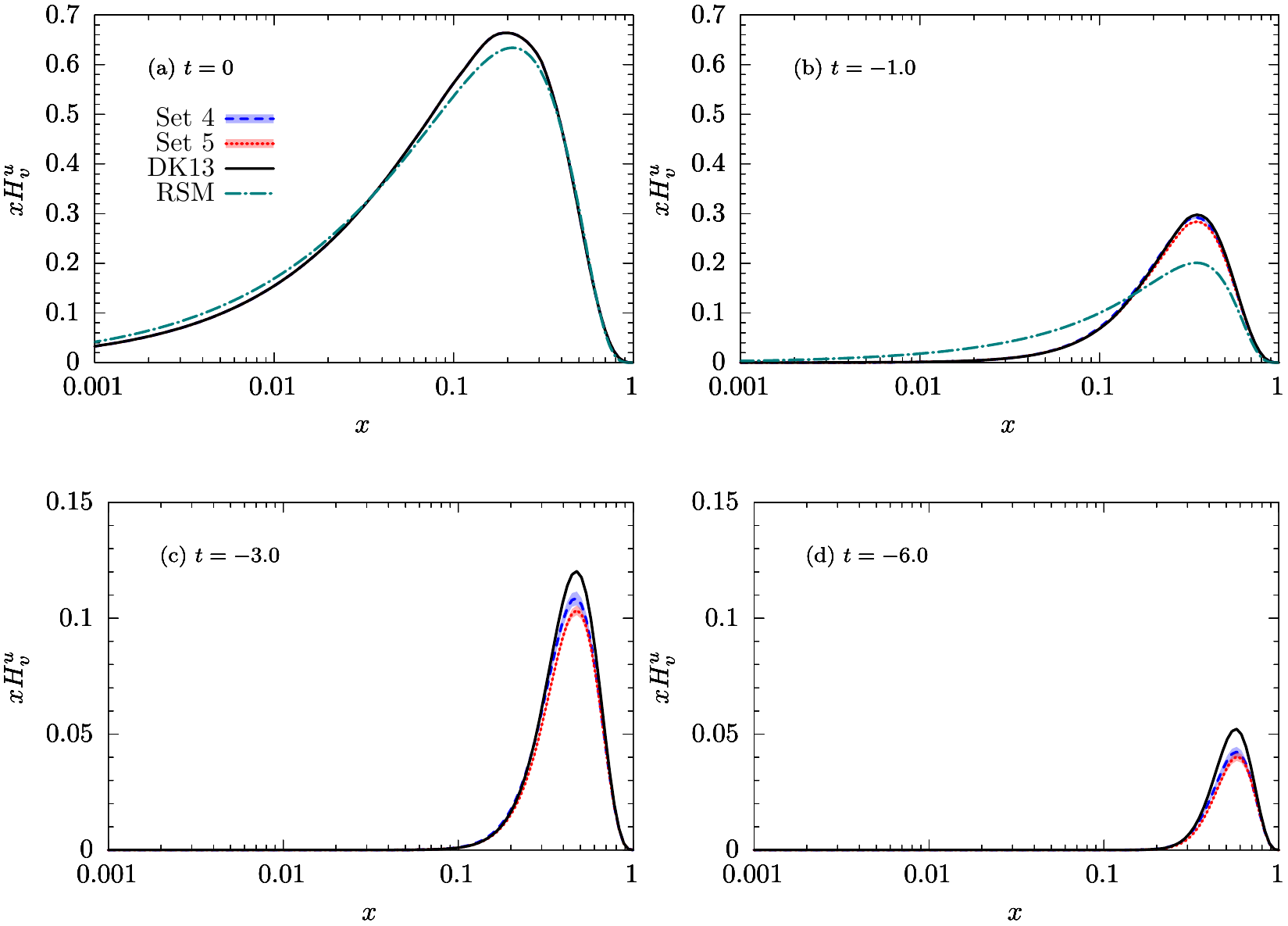}       
    \caption{A comparison between our results for GPD $ xH_v^u(x) $ obtained from two analyses
    	described in Sec. \ref{sec:four-two} and the corresponding ones from the analysis of
    	DK13 \cite{Diehl:2013xca} at four $ t $ values: (a) $ t=0$, (b) $t=-1$, (c) $t=-3$,
    	and (d) $t=-6 $ GeV$ ^2 $. The results of the RSM \cite{Kriesten:2021sqc} have
    	also been shown for $ t=0,-1$ GeV$ ^2 $.}
\label{fig:Hur}
\end{figure}

Fig.~\ref{fig:Hur} shows a comparison between our results for GPD $ xH_v^u(x) $ obtained from two analyses described in this subsection (Set 4 and Set 5) with their uncertainties and the corresponding ones from the analysis of \texttt{DK13} at four $ t $ values, $ t=0,-1,-3,-6 $ GeV$ ^2 $. Here, we have also included the recent results obtained from the Reggeized spectator model (RSM)~\cite{Kriesten:2021sqc}. Note that, as explained by the authors, their results are valid just for the values of $ -t $ less than unity. So, we have not plotted the corresponding ones for $ t=-3,-6 $ GeV$ ^2 $.
As can be seen, our results are very similar to \texttt{DK13}, though they are more suppressed with $ -t $ growing compared with \texttt{DK13}. Since the forward limit $ u_v(x) $ is the same for both cases as clearly seen from panel (a), the difference between our results and \texttt{DK13} at larger values of $ -t $ can be attributed to the difference in the profile functions $ f_{u_v}(x) $ of Eq.~(\ref{Eq4}). Although the RSM result is compatible with other groups at $ t=0 $, it becomes more different at larger values of $ -t $ as it is clear from panel (b).

\begin{figure}[!htb]
    \centering
\includegraphics{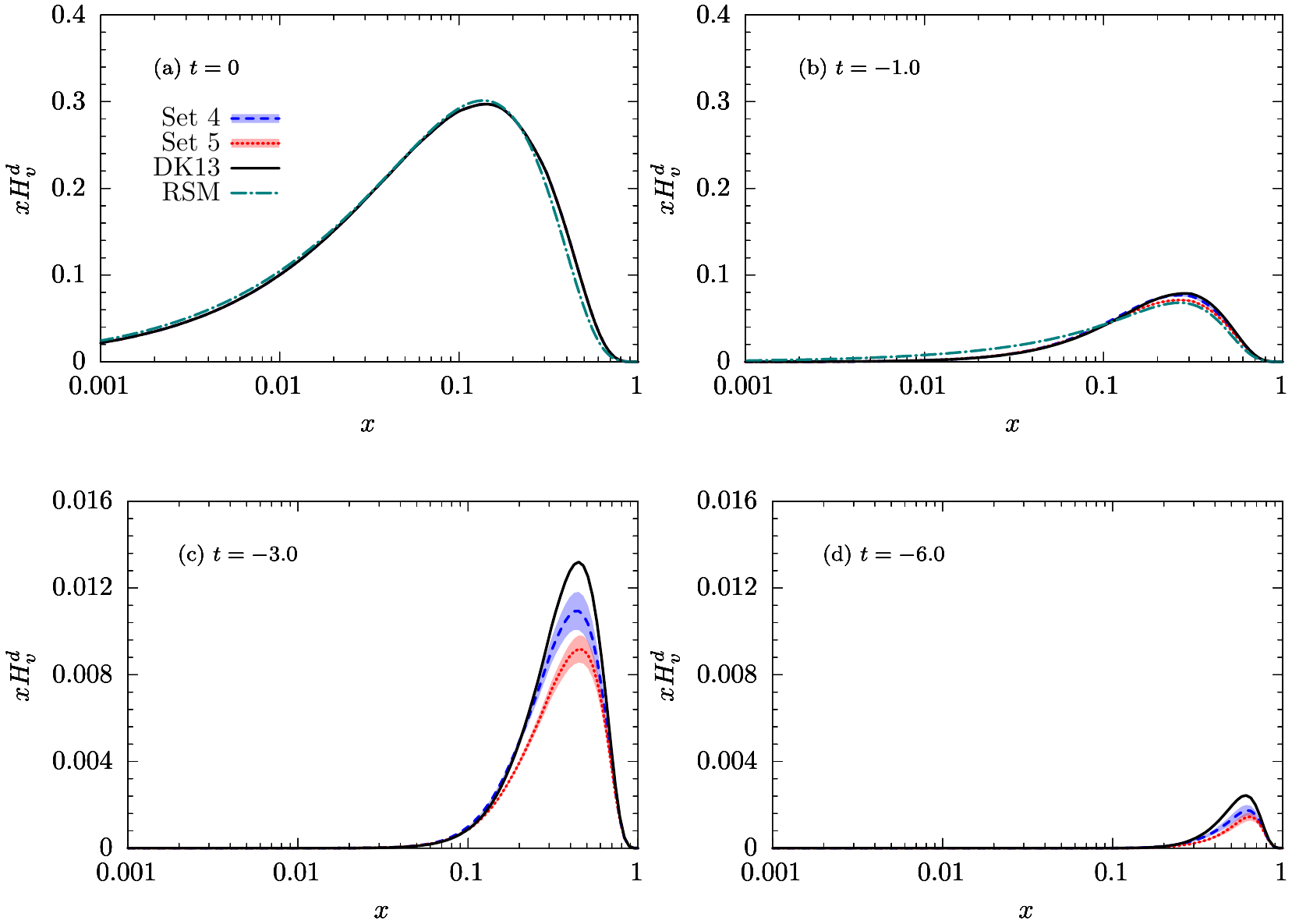}    
    \caption{Same as Fig.~\ref{fig:Hur} but for GPD $ xH_v^d(x) $. }
\label{fig:Hdr}
\end{figure}

Figure~\ref{fig:Hdr} shows the same results as Fig.~\ref{fig:Hur} but for GPD $ xH_v^d(x) $. In this case, our results become a little different rather than \texttt{DK13} with $ -t $ growing that indicates some differences in their profile function $ f_{d_v}(x) $. Note that \texttt{DK13} predicts again a larger distribution in magnitude compared with our results. Note also that in this case, the RSM prediction remains in good consistency with others even at larger values of $ -t $.

\begin{figure}[!htb]
    \centering
    \includegraphics{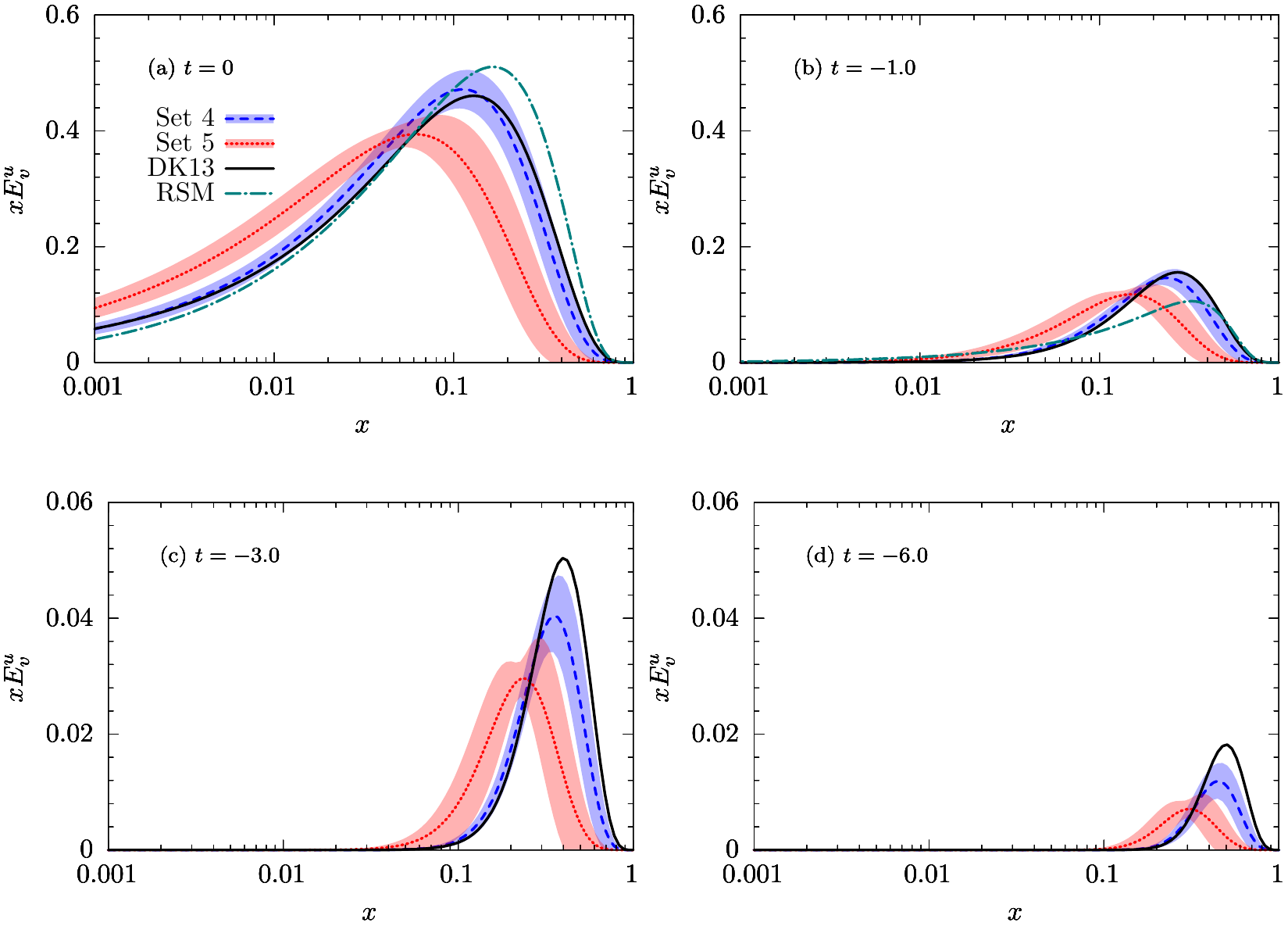}    
    \caption{Same as Fig.~\ref{fig:Hur} but for GPD $ xE_v^u(x) $. }
\label{fig:Eur}
\end{figure}

The results obtained for GPD $ xE_v^u(x) $ are shown in Fig.~\ref{fig:Eur} and compared again with the corresponding ones from \texttt{DK13} at $ t=0,-1,-3,-6 $ GeV$ ^2 $ as well as the RSM prediction. As can be seen, at $ t=0 $, Set 4 and \texttt{DK13} are almost similar, but they differ significantly with Set 5. This indicates that Set 4 and \texttt{DK13} have similar $ e_{v}^u(x) $ which tend to the larger values of $ x $ rather than $ e_{v}^u(x) $ of Set 5. However, they become also different by increasing the absolute value of $ t $ that indicates some differences in their profile function $ g_{u_v}(x) $ of Eq.~(\ref{Eq7}). The RSM result is more compatible with Set 4 and \texttt{DK13}, though its peak tends to the larger values of $ x $ with $ -t $ growing.

\begin{figure}[!htb]
    \centering
\includegraphics{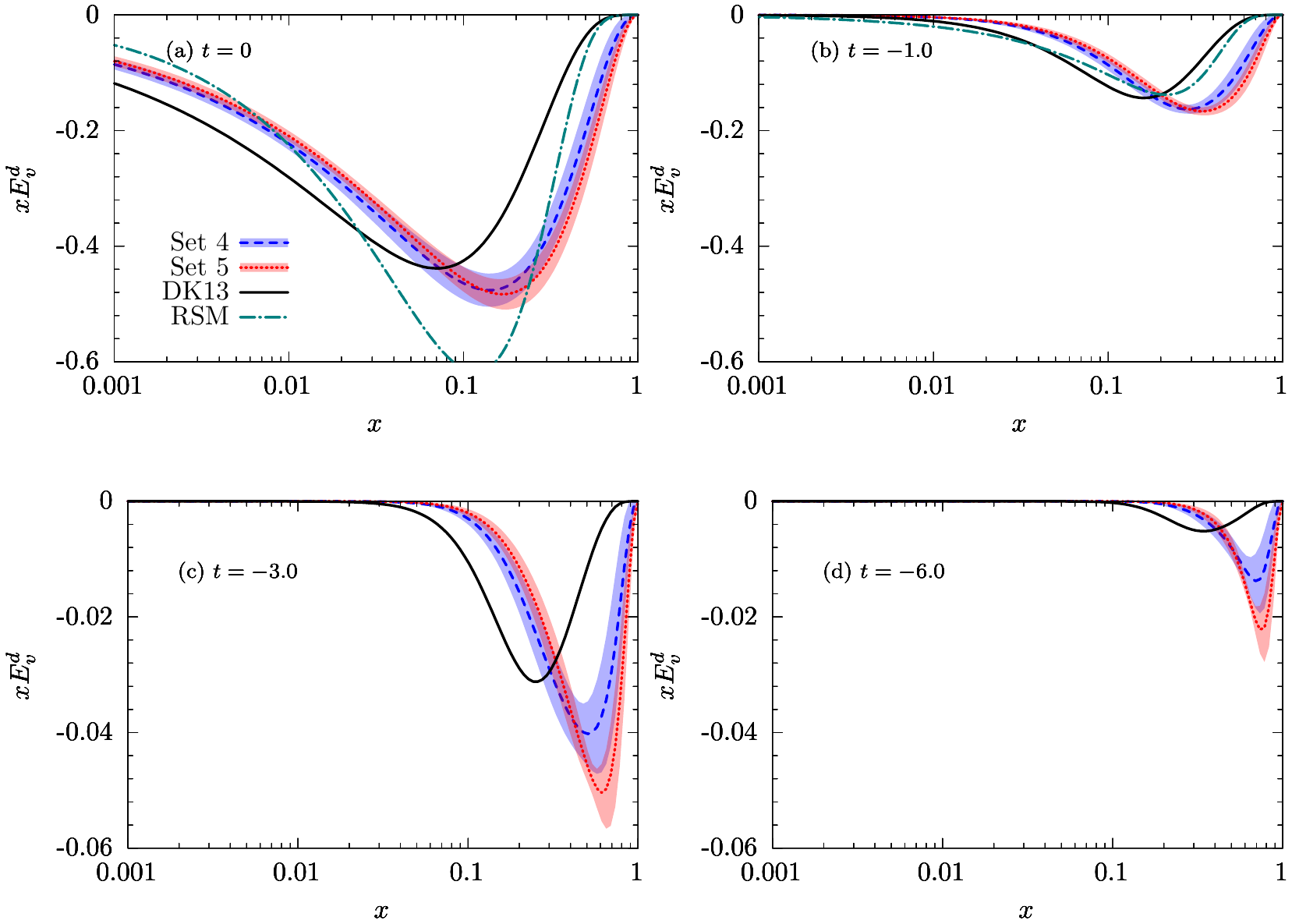}    
    \caption{Same as Fig.~\ref{fig:Hur} but for GPD $ xE_v^d(x) $. }
\label{fig:Edr}
\end{figure}

Figure~\ref{fig:Edr} shows the same results as Fig.~\ref{fig:Eur} but for GPD $ xE_v^d(x) $. In this case, our results are in good agreement with each other and some differences appear just with $ -t $ growing. However, they differ significantly from \texttt{DK13} at all values of $ -t $. Actually, our results tend to larger $ x $ values and have also larger magnitude compared with \texttt{DK13}. At $ t=0 $, the RSM behaves as our results at small $ x $ values, but it becomes different at medium and large values of $ x $. Although, the RSM and \texttt{DK13} are not in good consistency at $ t=0 $, they become similar at $ t=-1 $ GeV$ ^2 $.

Comparing the results obtained in this subsection with the corresponding ones from Sec.~\ref{sec:four-one}, one can conclude that the inclusion of the charge and magnetic radius data in the analysis put new constraints one GPDs, especially of $ E_v^q $. Actually, it is obvious from the results obtained that by inclusion of these data in the analysis, $x E_v^q $ distributions have significantly shifted to the large values of $ x $, while without these data they tend to localize at medium $ x $. Moreover, $ xH_v^q $ distributions are decreased in magnitude by inclusion the radius data. Note that, comparing with GPDs $ H_v^q $, there are still some differences between different sets of GPDs of $ E_v^q $. This indicates that it is necessary to include more precise experimental data in the analysis, especially those can put more constraints on the GPDs $ E_v^q $.

\subsection{Comparison with other quantities}\label{sec:four-three}

After extracting different sets of GPDs utilizing different scenarios and including different types of experimental data, it is interesting now to investigate how they describe the other physical quantities which are related to GPDs at zero skewness ($ \xi=0 $). 

It is well established now that $ n $th Mellin moment of the GPDs $ H $ and $ E $ can be expressed as polynomials in $ \xi $ of order $ n+1 $ considering the polynomiality property of the GPDs~\cite{Ji:1998pc}. For example, for the second Mellin moment of the GPD $ H $ we have~\cite{Polyakov:2002yz}
\begin{equation}
\int_{-1}^1 dx x \sum_q H^q(x,\xi,t) = M_2(t) + \frac{4}{5}\xi^2 d_1(t), 
\label{Eq18}
\end{equation}
where $ M_2(t) $ and $ d_1(t) $ are the gravitational FFs of the energy-momentum tensor. 
It is known that $ M_2(t) $ can provide information on the momentum fractions carried by the constituent quarks of the nucleon, while the information on the distribution of pressure and tensor forces
inside hadrons can be accessed from $ d_1(t) $ (called the D-term). For more information on gravitational FFs, one may refer to Refs.~\cite{Kobzarev:1962wt,Polyakov:2018zvc} as examples.

\begin{figure}[!htb]
    \centering
\includegraphics{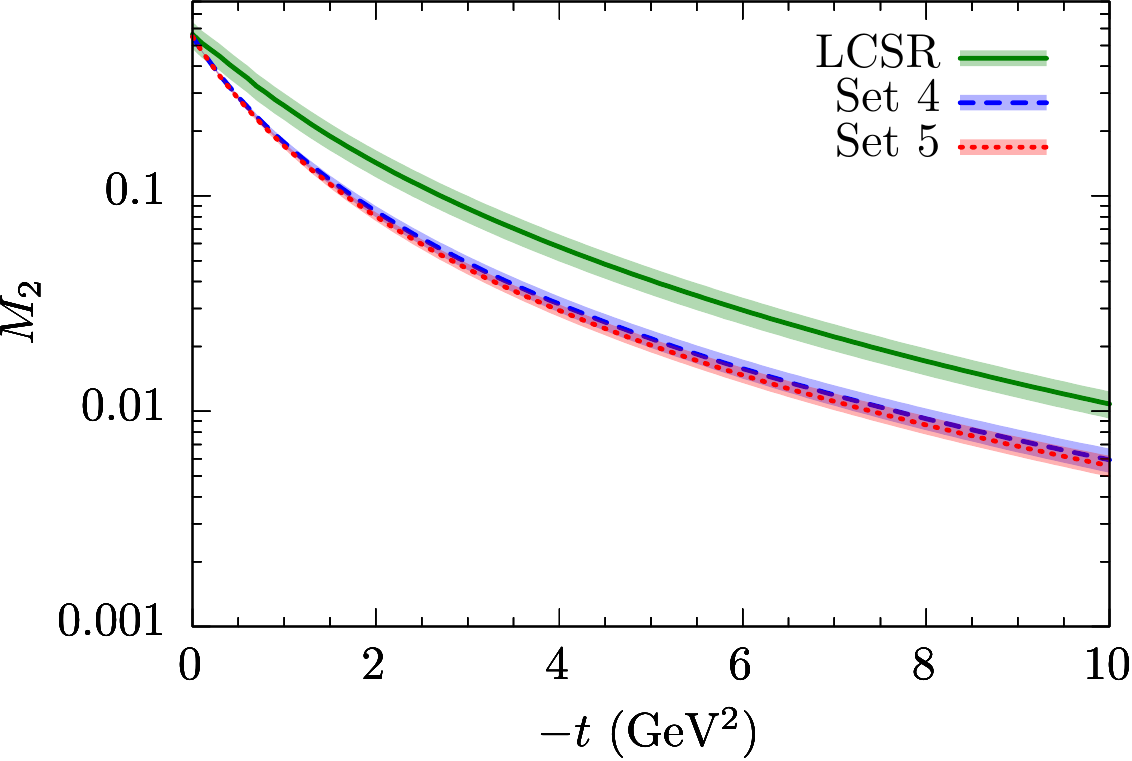}    
    \caption{A comparison between our results obtained for $ M_2(t) $ using Set 4 and Set 5 of GPDs and the corresponding one calculated using LCSR~\cite{Azizi:2019ytx}.}
\label{fig:M2t10}
\end{figure}

From Eq.~(\ref{Eq18}), it is obvious that $ M_2(t) $ is related to the GPDs $ H^q(x,t) $ at zero skewness.
Then, it may be an interesting idea to calculate $ M_2(t) $ using our GPDs obtained in Sec.~\ref{sec:four-two} as a function of $ -t $. Note that if both $ t $ and $ \xi $ are set to zero, the $ M_2 $ will be a simple sum of the momentum fractions carried by quarks, since GPDs $ H^q(x) $ turn into the ordinary PDFs $ q(x) $. Although there are no experimental measurements of $ M_2(t) $ unlike the D-term $ d_1(t) $, it is interesting to compare our results with the corresponding ones obtained from the light-cone QCD sum rules (LCSR)~\cite{Azizi:2019ytx}. Such a comparison has been shown in Fig.~\ref{fig:M2t10} where we have calculated the left hand of Eq.~(\ref{Eq18}) using our GPDs obtained in Sec.~\ref{sec:four-two} (Set 4 and Set 5) and compared them with the results of $ M_2(t) $ obtained using LCSR. It should be noted that since we have extracted only the valence GPDs from the experimental data and our results do not include the contributions from the sea quarks, we have considered an assumption to calculate $ M_2(t) $ that is using the profile functions of the valence quarks for the sea quarks too. Moreover, the LCSR result shown in Fig.~\ref{fig:M2t10} is the averaged results of LCSR presented in Ref.~\cite{Azizi:2019ytx} that has been evolved to $ \mu=2 $ GeV using the the renormalization group
equations extended to include mass renormalization (the original results are belonging to $ \mu=1 $ GeV).

\begin{figure}[!htb]
    \centering
\includegraphics{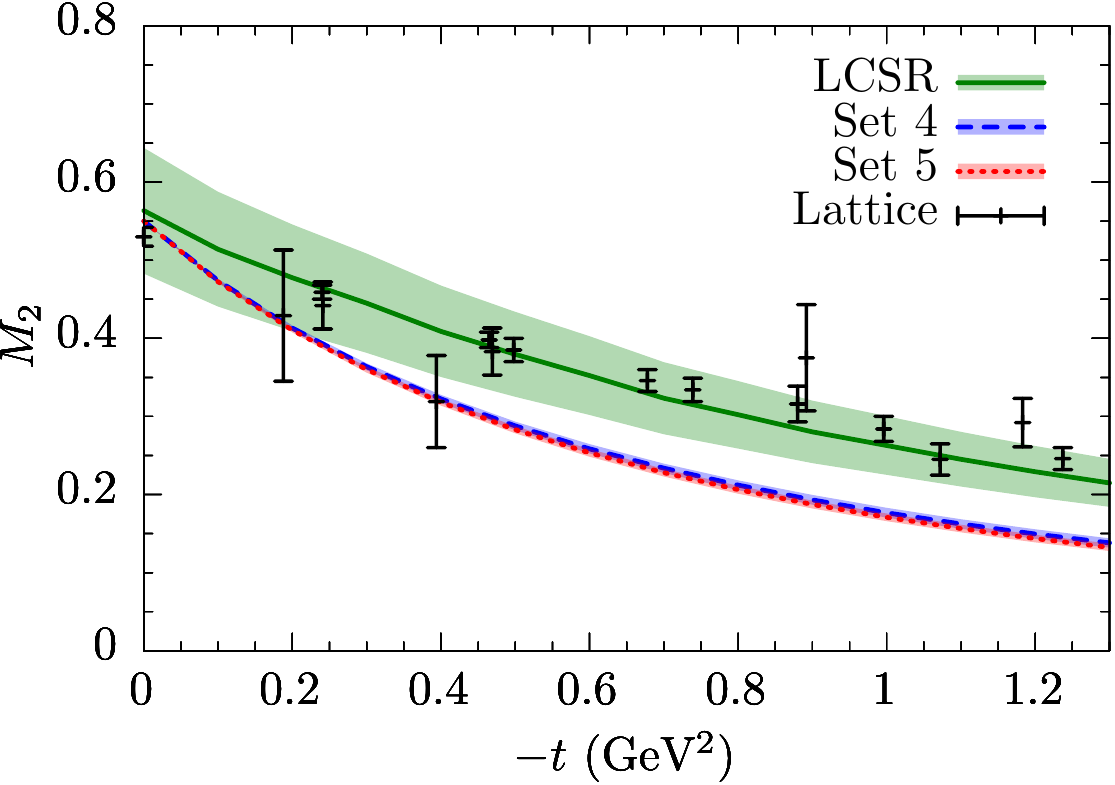}    
    \caption{Same as Fig.~\ref{fig:M2t10}, but in the interval $ 0 < -t < 1.3 $ GeV$ ^2 $ and including also the Lattice results taken from Table 15 of Ref.~\cite{LHPC:2007blg}.}
\label{fig:M2t1}
\end{figure}

As can be seen from Fig.~\ref{fig:M2t10}, the results are in good agreement with each other especially if one considers the uncertainties. Actually, it is very interesting that the result of a phenomenological approach in which GPDs are constrained from the experimental data is in a good consistency with the pure theoretical calculations. This indicates, on the other hand, the validity of the LCSR framework to study the hadrons structure and their properties. Note that the excellent agreement between Set 4 and Set 5 is a result of good constraints on GPDs $ H^q $. Figure~\ref{fig:M2t1} shows same comparison as Fig.~\ref{fig:M2t10} but in the interval $ 0 < -t < 1.3 $ GeV$ ^2 $ and including also the Lattice results taken from Table 15 of Ref.~\cite{LHPC:2007blg}. As can be seen, the consistency is also good in this case considering this fact that there is no explicit results from LCSR for $ -t < 1 $ GeV$ ^2 $ and the extrapolation has been used to extend the calculation to this interval.

\begin{figure}[!htb]
    \centering
\includegraphics{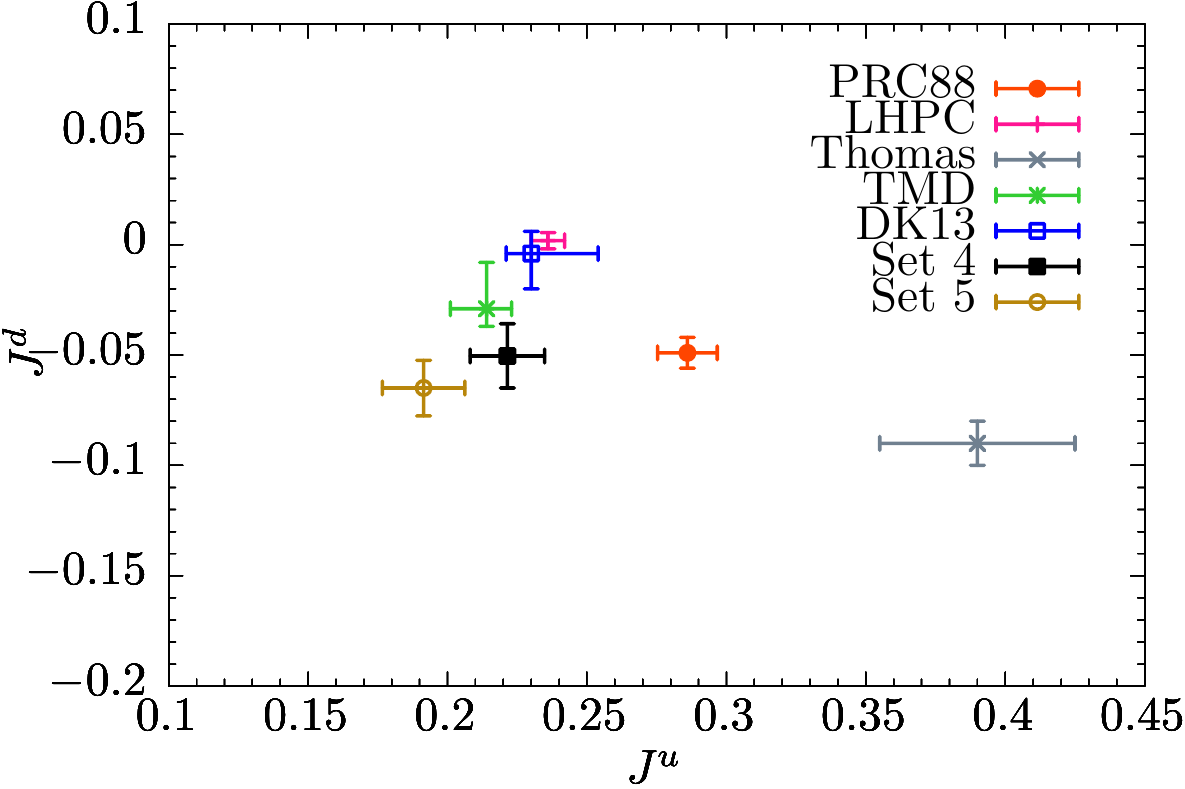}    
    \caption{A comparison between our predictions for $ J_v^u $ and $ J_v^d $ at limit $ t=0 $ and $ \mu=2 $ GeV obtained using Set 4 and Set 5 and the corresponding ones from PRC88~\cite{Gonzalez-Hernandez:2012xap}, LHPC~\cite{LHPC:2010jcs}, Thomas~\cite{Thomas:2008ga},  TMD~\cite{Bacchetta:2011gx}, and \texttt{DK13}~\cite{Diehl:2013xca}. }
\label{fig:JuJd}
\end{figure}

Another quantity that is related to GPDs is the total angular momentum carried
by the quarks inside the nucleon. In the most general case, it can be expressed in terms of GPDs $ H^q $ and $ E^q $ at zero skewness for each desired quark flavor $ q $ as
\begin{equation}
\int_{-1}^1 dx x [ H^q(x,t) +  E^q(x,t)] = \frac{1}{2} J^q(t).
\label{Eq19}
\end{equation}
It should be noted that this relation turns into the famous Ji's sum rule~\cite{Ji:1996ek,Goloskokov:2008ib} at $ t=0 $. Comparing with $ M_2 $, $ J^q $ may be more interesting for us since it contains also GPD $ E^q $. Since our analyses just includes the valence GPDs, we can calculate $ J_v^u $ and $ J_v^d $. Figure~\ref{fig:JuJd} shows a comparison between our predictions for $ J_v^u $ and $ J_v^d $ at limit $ t=0 $ and $ \mu=2 $ GeV obtained again using Set 4 and Set 5 and the corresponding ones from PRC88~\cite{Gonzalez-Hernandez:2012xap}, LHPC~\cite{LHPC:2010jcs}, Thomas~\cite{Thomas:2008ga},  TMD~\cite{Bacchetta:2011gx}, and \texttt{DK13}~\cite{Diehl:2013xca}. According to this figure, our results for $ J_v^u $ are in more agreement with LHPC, TMD, and \texttt{DK13} while they have significant difference with PRC88 and Thomas. For the case of $ J_v^d $, our results are in good agreement with PRC88 and TMD while differ significantly with other groups. Overall, one can conclude from Fig.~\ref{fig:JuJd} that our results, especially Set 4, are more consistent with LHPC, TMD and \texttt{DK13}. 

\begin{figure}[!htb]
    \centering
\includegraphics{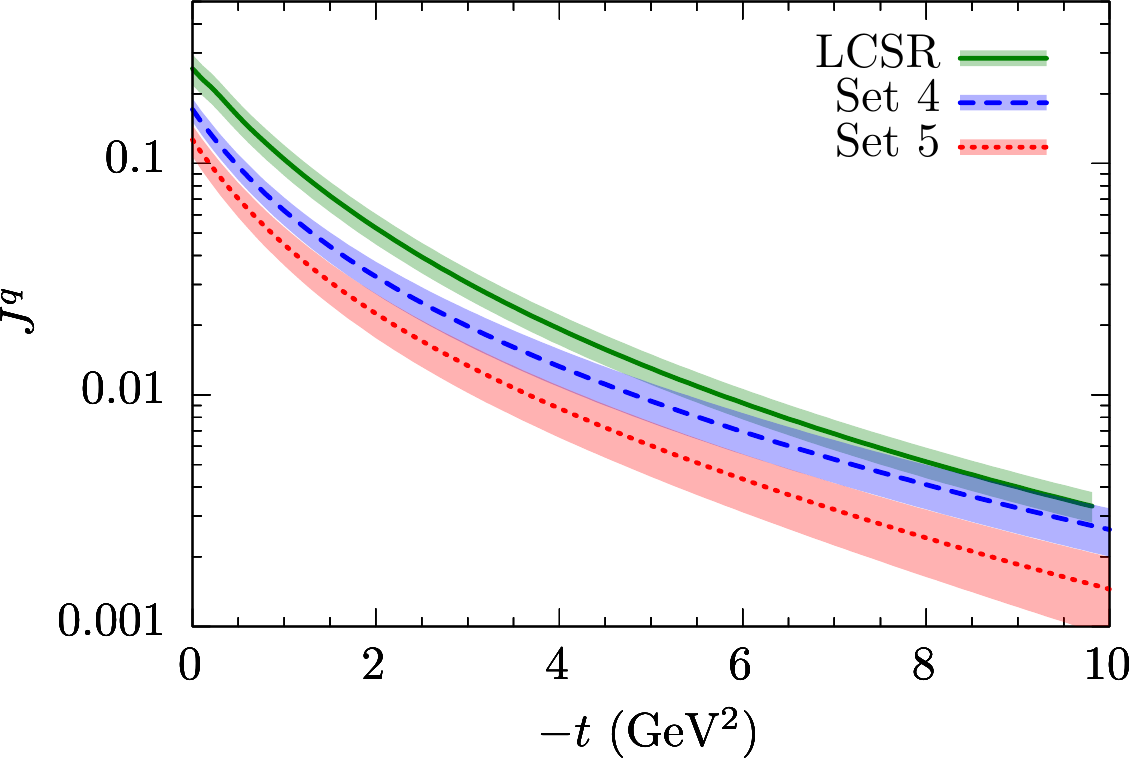}
    \caption{Same as Fig.~\ref{fig:M2t10}, but for the proton total angular momentum $ J^q $ of Eq.~(\ref{Eq19}).}
\label{fig:JFF}
\end{figure}

Other than Ji's sum rule for each specific flavor $ q $ at $ t=0 $, we can calculate Eq.~(\ref{Eq19}) as a function of $ t $ and also sum over all flavors to get the total angular momentum of the proton $ J^q $. However, since our analyses dose not include contributions from sea quarks we can just calculate $ J^q $ using the valence GPDs. Figure~\ref{fig:JFF} shows the results obtained using Set 4 and Set 5 and compares them with the corresponding ones from LCSR similar to Fig.~\ref{fig:M2t10} for $ M_2(t) $. Overall there is a good agreement between our results and LCSR again. However, the results are more different at small values of $ -t $ compared with Fig.~\ref{fig:M2t10}. Note that the difference between Set 4 and Set 5 in this case comes from the difference between GPDs $ E^q $ of these two sets. However, Set 4 is more consistent with LCSR. Moreover, some of the differences between our results and LCSR can be attributed to the absence of the sea quark contributions in our calculations.

\begin{figure}[!htb]
    \centering
\includegraphics{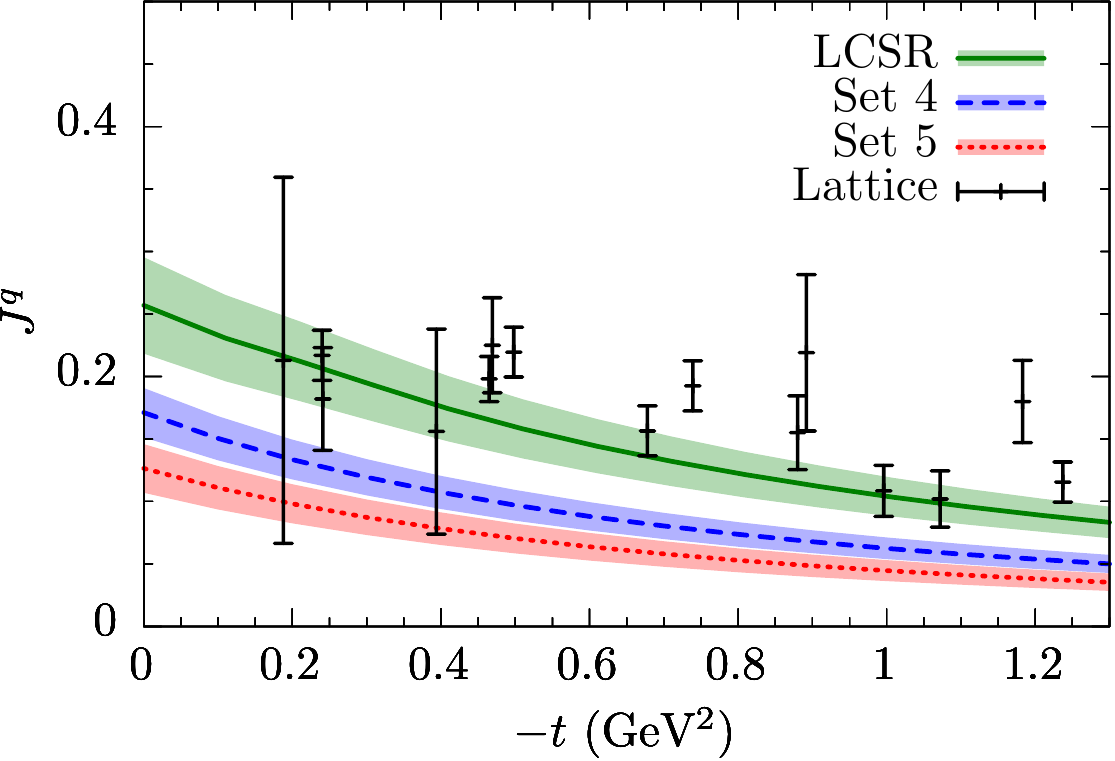}    
    \caption{Same as Fig.~\ref{fig:JFF}, but in the interval $ 0 < -t < 1.3 $ GeV$ ^2 $ and including also the lattice results taken from Table 15 of Ref.~\cite{LHPC:2007blg}.}
\label{fig:JqFF}
\end{figure}

We have compared the results obtained for $ J^q $ of the proton at the interval $ 0 < -t < 1.3 $ GeV$ ^2 $ in Fig.~\ref{fig:JqFF}. As before, the figure includes also the lattice results taken from Table 15 of Ref.~\cite{LHPC:2007blg}. Although the agreement between the results are acceptable considering uncertainties, but the differences are more considerable compared with Fig.~\ref{fig:M2t1} for $ M_2 $.
%

%

\section{Summary and conclusion}\label{sec:five} 

The 3D hadron structure can be accessed through GPDs~\cite{Muller:1994ses,Radyushkin:1996nd,Ji:1996nm,Ji:1996ek,Burkardt:2000za} which are measurable in hard exclusive scattering processes. In this work, following the recent works~\cite{Hashamipour:2019pgy,Hashamipour:2020kip} performed to determine the polarized GPDs $ \widetilde{H}_v^q $, we determined the unpolarized valence GPDs $ H_v^q $ and $ E_v^q $ with their uncertainties at zero skewness ($ \xi=0 $) by performing some $ \chi^2 $ analyses of the related experimental data. To this end, we first considered the world electron scattering data presented in Ref.~\cite{Ye:2017gyb} (\texttt{YAHL18}) where TPE corrections have also been incorporated. These data include the world $ R_p=\mu_p G_E^p/G_M^p $ polarization, world $ G_E^n $, and world $ G_M^n/\mu_n G_D $ data with overall 140 data points. They cover the  $ -t $ range from 0.00973 to 10 GeV$ ^{2} $. Utilizing two different scenarios where the optimum values of the parameters are obtained through a parametrization scan procedure~\cite{H1:2009pze}, we extracted three different sets of GPDs that all preserve the positivity condition and lead to an acceptable quality of the fit, especially for the neutron data. We compared our GPDs with the corresponding ones obtained from \texttt{DK13}~\cite{Diehl:2013xca} analysis. Although our results for $ H_v^q $ are in good consistencies with \texttt{DK13}, the results for $ E_v^q $ are different. Overall, we found that there are not enough constraints on GPDs $ E_v^q $ from FFs data solely though $ H_v^q $ are well constrained. 

As the next step, we included the data of the charge and magnetic radius of the nucleons in the analysis to investigate their impacts on the extracted GPDs. Utilizing two different scenarios again, we obtained two final sets of GPDs, namely Set 4 and Set 5, which are in more consistent with \texttt{DK13}, especially Set 4. We shown that the radius data put new constraints one the final GPDs, especially of $ E_v^q $. To be more precise, by inclusion of these data in the analysis, $ xH_v^q $ distributions are decreased in magnitude and $ xE_v^q $ are significantly shifted to the large values of $ x $ rather than before. 

As the final step, we calculated the gravitational FF $ M_2(t) $ and the total angular momentum of proton $ J^q(t) $ using our final sets of GPDs Set 4 and Set 5 and compared them with the results obtained from LCSR~\cite{Azizi:2019ytx} and Lattice QCD~\cite{LHPC:2007blg}. We shown that our results are interestingly in a good consistency with the pure theoretical predictions, especially Set 4. Overall, the differences are more considerable for the case of $ J^q(t) $ that includes also GPDs $ E_v^q $ compared to $ M_2(t) $. In addition, we calculated the total angular momentum carried by the valence quarks at $ t=0 $, namely Ji's sum rule~\cite{Ji:1996ek}, and compared our results with the corresponding ones obtained from other groups. Overall, our results, especially Set 4, are more consistent with LHPC~\cite{LHPC:2010jcs}, TMD~\cite{Bacchetta:2011gx} and \texttt{DK13}~\cite{Diehl:2013xca}. 

According to the results obtained in this paper, we emphasize that in order to get more universal GPDs, especially of $ E_v^q $, it is necessary to include more precise experimental data in the analysis.
In this regard, the future programs like one that will be done at Jefferson Lab~\cite{Accardi:2020swt} can shed new lights on the determination of GPDs.

%
\section*{ACKNOWLEDGMENTS}
M. Goharipour and K. Azizi are thankful to Iran Science Elites Federation (Saramadan) for the partial financial support provided under the grant No. ISEF/M/400150. 
M. Goharipour and H. Hashamipour thank the School of Particles and Accelerators, Institute for Research in Fundamental Sciences (IPM), for financial support provided for this research.

%


\end{document}